\documentclass{aa}
\usepackage[dvips]{graphicx}
\usepackage{psfig}
\usepackage{aabib99}
\begin{document}

   \thesaurus{2         
             (
              11.03.1; 
              11.03.4 MS1008--1224 
              11.04.1; 
              12.04.1; 
              12.07.1; 
              )}
 
   \title{Depletion curves in cluster lenses: simulations and application 
to the cluster MS1008--1224
   \thanks{Based on observations with the VLT--ANTU (UT1), operated on
   Cerro Paranal by the European Southern Observatory (ESO), Chile}}
   \titlerunning{Depletion curves of galaxy number counts}
   \author{Christophe Mayen \and Genevi\`eve Soucail}
   \offprints{C. Mayen, cmayen@ast.obs-mip.fr}

   \institute{ Observatoire Midi-Pyr\'en\'ees, Laboratoire
d'Astrophysique, UMR 5572, 14 Avenue E. Belin, F-31400 Toulouse, France
            }

   \date{Received ... ; Accepted ...}

   \maketitle

   \begin{abstract}
 
When the logarithmic slope of the galaxy counts is lower than 0.4 (this
is the case in all filters at large magnitude), the
magnification bias due to the lens makes the number density of objects
decrease, and consequently,
the radial distribution shows a typical depletion curve. 
In this paper, we present simulations of depletion curves obtained in
different filters for a variety of different lens models - e.g., a 
singular isothermal sphere, an isothermal sphere with a core, a 
power-law density profile, a singular isothermal ellipsoid and a 
NFW profile. 
The different model parameters give rise to different effects, and we show 
how the filters, model parameters and redshift distributions of
the background populations affect the depletion curves. We then
compare our simulations to deep VLT observations of the cluster
MS1008--1224 and propose to constrain the mass profile of the
cluster as well as the ellipticity and the orientation of the 
mass distribution. This application is based solely on deep photometry of the
field and does not require the measurement of shape parameters of 
faint background galaxies. We finally give some possible applications
of this method, useful for cluster lenses. 

   \end{abstract}


\section{Introduction}

Since a few years, a new application of gravitational lensing in
clusters of galaxies has started to be explored, namely the depletion
effect of number counts of background galaxies in cluster centers.
This effect results from the competition between the gravitational
magnification that increases the detection of individual objects 
(at least for flux limited data and marginally resolved objects) and
the deviation of light beam that spatially magnifies the observed area
and thus decreases the apparent number density of sources. 

This effect was pointed out as a possible application of the
magnification bias by 
Broadhurst et al. \cite*{broadhurst2} where they suggested a new method for measuring
the projected mass distribution of galaxy clusters, based solely on
gravitational magnification of background populations by the
cluster gravitational potential. In addition, they suggested that the
mass-sheet degeneracy, initially pointed out by
Schneider and Seitz \cite*{schneiderseitz} and observed in mass reconstruction with weak
lensing measures, could be broken by using gravitational
magnification information which is directly provided by the depletion
curves. This method has been used by Fort et al. \cite*{fort} and  
by Taylor et al. \cite*{taylor} 
to reconstruct a two-dimensional mass map of Abell 1689
in the innermost 27 arcmin$^2$, taking into account the nonlinear
clustering of the background population and shot noise. The results
are consistent with those inferred from weak shear measurements and
from strong lensing. However, the surface mass density cannot be
obtained from magnification alone since magnification also depends
on the shear caused by matter outside the data field 
\cite{young,schneiderweaklensing}. But in practice, if 
the data field is
sufficiently large and no mass concentration lies close to but outside
the data field, the mass reconstruction obtained from magnification
can be quite accurate \cite{schneiderking}.

This method is an attractive alternative to weak lensing 
because it is only based on galaxy counts and does not require the
measure of shape parameters of very faint galaxies.  In addition,
it is still valid in the intermediate lensing regime, close to the
cluster center, without any strong modifications of the
formalism. Meanwhile, it is more sensitive to Poisson noise which
increases when the number density decreases in the depletion
area. Another weak point identified by Schneider et al. \cite*{schneiderking} is that it may
significantly depend on the galaxy clustering of background
objects which can have large fluctuations from one cluster to another.

A second application of the magnification bias that has been suggested
is to use the shape and the width of depletion curves to reveal the
redshift distribution of the background populations. This technique
was first used by Fort et al. \cite*{fort} in the cluster Cl0024+1654 to study the
redshift distribution of background sources in the range
$26<\mathrm{B}<28$ and $24<\mathrm{I}< 26.5$. They found that $60\%
\pm 10\%$ of the population in the B band is located between $z=0.9$
and $z=1.1$ while the remaining galaxies are broadly distributed
around $z=3$. The population in the I band shows a similar redshift
distribution but it extends to larger redshift $(\simeq 20\%$ of
objects at $z>4)$. 

The last application of the magnification bias that has been explored
is the search for constraints on cosmological parameters. This method
is based on the fact that the ratio of the two extreme radii which
delimit the depletion area depends on the ratio of the angular
distances lens-source and observer-source. Consequently it depends on
the cosmological parameters, as soon as all the redshifts are fixed,
and mainly on the cosmological constant. This method was first used by Fort et al. 
\cite*{fort} in Cl0024+1654 and in Abell 370. Their results favor a
flat cosmology with $0.6<\Omega_{\Lambda}<0.85$ and are consistent
with those obtained from other independent methods (see White \cite*{white}). This 
technique requires a good modeling of the
lens and could be improved by being extended to a large number of
cluster lenses and by using an independent estimate of the redshift
distribution (for example from photometric redshifts). However, this method 
was disputed by Asada \cite*{asada} who claimed that it is difficult to 
determine $\Omega$ by this method without the assumption of a spatially 
flat universe. He also mentioned the uncertainty of the lens model as 
one of the most serious problems in this test and concluded that this method 
cannot be taken as a clear cosmological test to determine $\Lambda$.

In order to better understand the formation of depletion curves and
their dependence with the main characteristics of the lenses and the
sources, we developed a detailed modeling of the curves in various
conditions.  In Sect. 2 we present the lens and counts models
used in our simulations. In Sect. 3 we study the influence of the
lens mass profile, with several sets of analytical mass
distributions. In Sect. 4 we explore the influence of background
sources distribution, selected through different filters. An
application on real data is presented in Sect. 5, in the cluster MS
1008--1224, observed with the VLT.  We compare these observations with
the simulations in order to derive some constraints on the mass profile
and the ellipticity of the cluster, as well as on the background
sources distribution. Some prospects about the future of this method
are given as a conclusion in Sect. 6.\\ Throughout the paper, we
adopt a Hubble constant of H$_0=50$ km s$^{-1}$ Mpc$^{-1}$, with
$\Omega_\Lambda=0$ and $\Omega_0 = 1$.


\section{Modeling depletion curves}

\subsection{The magnification bias}
\label{mu}
The projected number density of objects magnified by a factor $\mu$
through a lensing cluster and with magnitude smaller than $m$ can be
written in the standard form:
\[
\mathrm{N}(<m) = \mathrm{N}_{0}(<m) \ \mu ^{2.5\gamma -1}
\]
where $\mathrm{N}_{0}$ is the number density of objects in an empty
field and $\gamma$ is the logarithmic slope of the galaxy number
counts. Note first that this relation applies only for background
galaxies. This means that we implicitly work at faint magnitude where
the foreground counts are reduced and do not affect the slope
$\gamma$. Note also that the magnification $\mu$ depends on the source
redshift through the geometric factor $\mathrm{D}_{LS} /
\mathrm{D}_{OS}$ in a significant way. For example, for $z_L = 0.4$,
$\mathrm{D}_{LS} / \mathrm{D}_{OS}$ changes by more than 40\%\ when
$z_S$ varies from 0.8 to 2.  For a lower redshift cluster lens, the
effect should be reduced, at least for all sources at $z > 0.8$.

An exact writing of the magnification bias, which also
takes into account the local effect of magnification, changing with
radius $r$ within the lens can be re-formulated as:
\begin{eqnarray}
& & \frac{\mathrm{N}(r, <m)}{\mathrm{N}_{0}(<m)} = \\
& & = \frac {\int_0^{z_L} n(z, <m) \ dz + 
\int_{z_L}^{+\infty} n(z, <m) \ \mu(r,z)^{2.5 \gamma -1} \ dz}
{\int_{0}^{+\infty} n(z, <m) \ dz} \nonumber \\
& & = 1 - \frac {\int_{z_L}^{+\infty} n(z, <m) 
\left[ 1 - \mu(r,z)^{2.5 \gamma -1} \right] dz} 
{\mathrm{N}_{0}(<m)} \nonumber 
\end{eqnarray}
where $n(z, <m)$  is the density of galaxies of apparent magnitude
smaller than $m$, located at redshift $z$. $\gamma(z)$ represents the
shape of the luminosity function at redshift $z$ and we will assume in
the following that $\gamma$ does not depends on redshift, {\em i.e.}
most of the faint galaxies are representative of the power-law part of
the luminosity function. In addition we suppose that the slope is a
constant with redshift. 

It is clear that the radial behavior of the ratio $\mathrm{N}(r, <m)
/ \mathrm{N}_{0}(<m)$ is strongly related to $\mu (r)$. It decreases
from the center, up to a minimum at $r \sim R_E$ when magnification
goes to infinity (Einstein radius) and then increases up to 1 again,
when magnification goes to 0. This is the so-called radial depletion
curve which has been detected in a few clusters already 
\cite{fort,taylor,athreya}.

From the lensing point of view, the magnification $\mu$ can be written
as \cite{lenses} :
\begin{equation}
\mu=\frac{1}{\mathrm{det}\mathcal{A}}=\frac{1}{(1-\kappa)^{2}-\gamma^{2}}
\end{equation}
where $\mathcal{A}$ is the magnification matrix. The convergence
$\kappa$ and the shear $\gamma$ are expressed in cartesian
coordinates as a function of the reduced gravitational potential
$\varphi$ by :
\begin{equation}
\kappa=\frac{1}{2}\left( \frac{\partial^{2}\varphi}{\partial
x^2}+\frac{\partial^{2}\varphi}{\partial y^2}\right)
\end{equation}
\begin{equation}
\gamma=\sqrt{\gamma_1^{2}+\gamma_2^{2}}
\end{equation}
with :
\[
\gamma_1=\frac{1}{2}\left( \frac{\partial^{2}\varphi}{\partial
x^2}-\frac{\partial^{2}\varphi}{\partial y^2}\right) 
\quad {\mathrm{and}} \quad
\gamma_2=\frac{\partial^{2}\varphi}{\partial x \partial y}
\]
All these factors depend on the lens and source redshifts, as
$\varphi$ is related to the true projected potential $\Phi$ by :
\[ \varphi = \frac{2}{c^2} \ \frac{1}{\zeta^2} \ \frac{\mathrm{D}_{LS}
\mathrm{D}_{OL}} {\mathrm{D}_{OS}} \ \Phi  \]
where $\zeta$ is a characteristic scale of the lens. 

In the weak lensing regime we have the approximation $\mu(r) \sim 1 +
2 \kappa(r)$ \cite{broadhurst2}. So the potential interest 
of the depletion curves is
that they trace directly the convergence distribution, or equivalently
the surface mass density distribution. In principle, they may allow an
easy mass reconstruction of lenses in the weak regime. We will see
below what are the main limitations of such curves, already explored
by Athreya et al. \cite*{athreya}.

\subsection{The cluster lens}
Different sets of models are used, with increasing complexity. In all
cases we suppose that the lens redshift is 0.4, with reference to
the cluster Cl0024+1654 studied by Fort et al. \cite*{fort}. With the adopted
cosmology this means that the scaling corresponds to 6.4 $h_{50}^{-1}$
kpc for 1\arcsec . For each model we will use an analytic expression
and develop it to compute analytically the magnification and its
dependence with redshift and radius. All the models are scaled in
terms of the Einstein radius $R_E$. 

\subsubsection{The singular isothermal sphere (SIS)}
This model is the simplest one which can describe a cluster of
galaxies and is very useful for the analytical calculations of the
gravitational magnification. However it has also some physical meanings:
\begin{itemize}
\item[$\bullet$] it corresponds to a solution of the Jeans equation
and thus it can be written as a function of the observed velocity
dispersion;
\item[$\bullet$] in a non-collisionnal description of the collapse of a
self-gravitating system, the violent relaxation, during which particles
exchange energy with the average field, leads to an isothermal
distribution \cite{binney}. 
\end{itemize} 
But, this model is only valid inside certain limits due to the
divergence of the central mass density and of the total mass to
infinity. This divergence has no consequences on our work because the
validity limits of the model correspond to the limits of the depletion
regime. 
The density of matter can be written as :
\begin{equation}
\rho (r) =\rho_{E}\ \left(\frac{r}{R_E}\right)^{-2} 
\end{equation}
where $R_E$ is the Einstein radius :
\[
R_{E}=4\pi \ \frac{\sigma ^2}{c^2} \ \frac{D_{LS} D_{OL}}{D_{OS}}
\]
and $\sigma$ is the velocity dispersion along the line of sight. 
The gravitational magnification is simply :
\begin{equation}
\mu^{-1} = 1-\frac{R_E}{r} 
\end{equation}

\subsubsection{The isothermal sphere with core radius} This model
avoids the divergence of the mass density in the inner part of the
cluster with an internal cut-off of the density distribution.  We
chose the following distribution, as described in Hinshaw and Krauss \cite*{hinshaw} 
or Grossman and Saha \cite*{grossman}. The density of matter is given by :
\begin{equation}
\rho (r) = \frac{3}{2 \pi} \ \frac{\Phi_{0}}{G R_{C}^{2}} \ 
\frac{1+\frac{x^{2}}{3}\left|\Delta^{2}-1\right|}{(1+x^2
\left|\Delta^{2}-1\right|)^2}
\end{equation}
where $R_C$ is the core radius, $x=r/R_{E}$ and $\Phi_{0}$ is the
central value of the gravitational potential.

In this case, the Einstein radius is $R_{E} = R_{C} \sqrt{\Delta^{2}-1}$, 
where : 
\[
\Delta = \frac{4\pi \sigma^{2}}{c^{2}} \ 
\frac{D_{LS}}{D_{OS}} \ \frac{D_{OL}}{R_{C}}
\]
The magnification can be written as :
\begin{eqnarray}
\mu^{-1} = \left[ 1-\Delta\left(1+x^{2}\left|\Delta^{2}-1\right|\right)
^{-\frac{3}{2}}\right] \times  \\ 
\qquad \qquad \left[ 1-\Delta\left(1+x^{2}\left|\Delta^{2}-1\right|\right)^
{-\frac{1}{2}}\right] \nonumber
\end{eqnarray}

Physically, in most cases the core radius is smaller than or comparable to
the Einstein radius, so we do not expect strong effects in the outer
parts of the depletion curves.

\subsubsection{A power-law density profile} 
This model is a generalization of the SIS. It
can be used in order to test the departure from an isothermal profile
far away from the cluster center. The density is given by :
\begin{equation}
\rho = \rho_{E} \ \left(\frac{r}{R_E}\right)^{-\alpha} 
\end{equation}
where $\alpha$ is the logarithmic slope of the density profile. In order
to
keep a physical model for the mass distribution, we must have  
$\alpha <3$. The Einstein radius can be written as :
\[
R_E=\sqrt{\frac{4}{c^2} \ \frac{D_{LS} D_{OL}}{D_{OS}} \ GM_E \
I_{\alpha/2}}
\]
where $M_{E} = 4\pi\rho_{E} R_{E}^{3} / \left( 3-\alpha \right)$ is
the integrated mass inside the Einstein radius and
$I_{\beta}=\int_{0}^{+\infty} (1+\xi^2)^{-\beta} \, d\xi$.
The magnification is then :
\begin{equation}
\mu^{-1} = \left[ 1-(2-\alpha) \left(\frac{r}{R_E}\right)^{1-\alpha} \right] 
\times \left[ 1- \left(\frac{r}{R_E}\right)^{1-\alpha} \right] 
\end{equation}

\subsubsection{The singular isothermal ellipsoid} 
This type of model is interesting as a lot of clusters have elliptical
shapes in their galaxy distribution or their X-ray isophotes
\cite{buote1,buote2,lewis,soucail}. Thus, an elliptical lens  
represents a more realistic model although it is still reasonably
simple. In order to study the effect of the ellipticity of the
potential $\epsilon=1-\frac{b}{a}\; (a\geq b)$ on the depletion, we
introduced the singular isothermal ellipsoid in our simulations
\cite{kormann}.  Using polar coordinates $(r,\varphi)$ in the lens
plane we introduce
\begin{equation}
\zeta = \sqrt{\left(r\cos \varphi\right)^2 + \left(1-\epsilon \right)^2
\left(r\sin \varphi\right)^2} 
\end{equation}
which is constant on ellipses with minor axis $\zeta$ and major axis
$\zeta/\left(1-\epsilon \right)$. The surface mass density can then be
written as :
\begin{equation}
\Sigma \left(r,\varphi \right) = \frac{\sqrt{1-\epsilon} \: \sigma^2}{2G}
\frac{1}{\zeta}
\end{equation}
where $\sigma$ is the velocity dispersion along the line of sight.\\
The convergence is :
\begin{equation}
\kappa = \frac{\Sigma \left(r,\varphi \right)}{\Sigma_{critical}} = 
\Sigma \left(r,\varphi \right) \times \frac{4\pi G}{c^2} \ 
\frac{D_{OL} D_{LS}}{D_{OS}} 
\end{equation}
The magnification writes then quite simply:
\begin{equation}
\mu^{-1} = 1 - 2 \kappa 
\end{equation}

\subsubsection{NFW profile}
We introduced in our simulations the universal density profile of Navarro et al. 
\cite*{navarro} for dark matter halos which was found from
cosmological simulations of the growth of massive structures. This
profile is a very good description of the radial mass distribution
inside the virial radius $r_{200}$. Wright and Brainerd \cite*{wright} have compared it
with a SIS for several cosmological models. They find that the
assumption of an isothermal sphere potential results in an
overestimate of the halos mass which increases linearly with the value
of the NFW concentration parameter.  This overestimate depends upon
the cosmology and is smaller for rich clusters than for galaxy-sized
halos.\\ 
The NFW density profile is given by :
\begin{equation}
\rho (r) = \frac{\delta_c
\rho_c}{\left(r/r_s\right)\left(1+r/r_s\right)^2}
\end{equation}
where $r_s = r_{200}/c$ is a characteristic radius and $\rho_c$
is the critical density. $c$ is the concentration parameter and 
$\delta_c$ is a characteristic over-density for the halo which is related
to $c$ by the requirement that the mean density within the virial radius
should be $200 \rho_c$. This leads to : 
\[
\delta_c = \frac{200}{3}\frac{c^3}{\ln (1+c)-c/(1+c)}
\]
If we take $x = r / r_s$, the surface mass density and the shear of
a NFW lens can be written as \cite{wright}: 
\[
\kappa_{\mathrm{NFW}}(x)=\frac{r_s \delta_c \rho_c}{\Sigma_c}
\left\lbrace
\begin{array}{lr}
f_{<}(x) & (x<1) \\
2/3 & \qquad (x=1) \\
f_{>}(x) & (x>1) 
\end{array}
\right. 
\]

\[
\gamma_{\mathrm{NFW}}(x)=\frac{r_s \delta_c \rho_c}{\Sigma_c}
\left\lbrace
\begin{array}{lr}
g_{<}(x) & (x<1) \\
10/3 + 4\ln (1/2) & \quad (x=1) \\
g_{>}(x) & (x>1)
\end{array}
\right. 
\]
where $\Sigma_c$ is the critical surface mass density.
$f_{<}(x)$ and $f_{>}(x)$ express as :
\[
f_{<}(x) = \frac{2}{\left(x^2-1\right)}
\left( 1- \frac{2}{\sqrt{1-x^2}}
\arg \mathrm{th}\sqrt{\frac{1-x}{1+x}}\,\right)
\]
\[
f_{>}(x) = \frac{2}{\left(x^2-1\right)}
\left( 1- \frac{2}{\sqrt{x^2-1}}
\arctan\sqrt{\frac{x-1}{x+1}}\,\right) 
\]
while $g_{<}(x)$ and $g_{>}(x)$ express as :
\begin{eqnarray}
g_{<}(x) & = & \frac{8 \arg \mathrm{th}\sqrt{\frac{1-x}{1+x}}}{x^2\sqrt{1-x^2}}
+\frac{4}{x^2}\ln \left(\frac{x}{2}\right) \nonumber \\ 
& & \qquad -\frac{2}{\left(x^2-1\right)}+\frac{4
\arg \mathrm{th}\sqrt{\frac{1-x}{1+x}}}{\left(x^2-1\right)\sqrt{1-x^2}}
\nonumber
\end{eqnarray}
\begin{eqnarray}
g_{>}(x) & = & \frac{8 \arctan \sqrt{\frac{x-1}{x+1}}}{x^2\sqrt{x^2-1}}
+\frac{4}{x^2}\ln \left(\frac{x}{2}\right) \nonumber \\ 
& & \qquad -\frac{2}{\left(x^2-1\right)}+\frac{4
\arctan \sqrt{\frac{x-1}{x+1}}}{\left(x^2-1\right)^{\frac{3}{2}}}
\nonumber
\end{eqnarray}
The magnification by the NFW lens is then :
\begin{equation}
\mu^{-1} = \left(1-\kappa_{\mathrm{NFW}}\right)^2 -
\gamma_{\mathrm{NFW}}^2
\end{equation}

\subsection{The galaxy redshift distribution}
\subsubsection{Analytical distribution}

We used the analytical redshift distribution introduced by Taylor et al. 
\cite*{taylor} to fit the redshift distribution of the galaxies in the
range $20<I<24$ for $0.25<z<1.5$ :
\begin{equation}
N(z, I<24)=N_0 (I<24) \frac{\alpha
z^2}{z_{*}^{3}\Gamma(3/\alpha)}\exp\left[-\left(
\frac{z}{z_*}\right)^{\alpha}\right]
\end{equation}
with $\alpha=1.8$ and $z_*=0.78$. Here, we extend the redshift range
between $z=0.05$ and $z=3$ and we take $N_0 (I<24) = 12$ objects per
arcmin$^2$ which is the count rate found by Taylor et al. \cite*{taylor} for red
galaxies with $20<I<24$.  The distribution shows a maximum at $z=0.8$
and has an average redshift of $< z > \simeq 1$. It is worth noting
that with these parameters, the redshift distribution does not
correspond to very deep observations.

\subsubsection{The magnitude-redshift distributions in the U to K
photometric bands}
\label{countmodel}
We also used a model of galaxy number counts from which the
magnitude-redshift distribution in empty field can be computed with a
large set of photometric bands. This model was developed by B\'ezecourt et al. \cite*{jocelyn2} 
and largely inspired by Pozetti et al. \cite*{pozetti}. It includes
the model of galaxy evolution developed by Bruzual and Charlot \cite*{bruzual93}, with
the upgraded version so-called GISSEL, and standard parameters for the
initial mass function (IMF) and star formation rates (SFR) for
different galaxy types. In order to reproduce the deep number counts
of galaxies, evolution of the number density of galaxies is included
to compensate for the smaller volumes in an $\Omega_0 = 1$ universe,
following the prescriptions of Rocca-Volmerange and Guiderdoni \cite*{rocca}. This model reproduces
fairly well deep number counts up to $B=27$ or $I=25$, as well as
redshift distributions of galaxies up to $B=24$ or $I=22$
\cite{jocelyn2}.

\begin{table}
\caption{Magnitude ranges adopted for the field number counts in the
simulations (see text for more details concerning the number count model).}
\label{tab1}
\begin{center}
\begin{tabular}{ccc}
\hline\noalign{\smallskip}
Filter &  Magnitude range & Faint end slope \\
\noalign{\smallskip}\hline\noalign{\smallskip}
B  & 22 -- 28 & 0.14  \\
I  & 20 -- 26 & 0.23  \\
K  & 18 -- 23 & 0.22  \\
\noalign{\smallskip}\hline
\end{tabular}
\end{center}
\end{table}
The advantage of this model is that the galaxy distribution can be
computed for any photometric band, and the effects of the color
distribution of sources can be explored. In practice we will limit our
study in this paper to B, I and K bands (Table \ref{tab1}). 


\section{Influence of the lens parameters on depletion curves} 
For our first set of simulations, we used the analytical redshift
distribution for the sources, allowing a fully analytical treatment of
the simulations. The contamination by foreground objects with this
distribution is weak (about $\simeq 9 \%$ of foreground galaxies for
$z_L=0.4$).  In each case, we computed depletion curves, and then
examined their behavior. Note that we limited our analysis outside
the first 20\arcsec\ from the center where the signal cannot be
constrained observationally (decrease in the observed area for each
point, obscuration by the brightest cluster galaxies \ldots). For these
reasons, this ``forbidden'' area will be shaded in each plot.

\subsection{Influence of the velocity dispersion}
\begin{figure}
\centerline{
\resizebox{\hsize}{!}{
\includegraphics{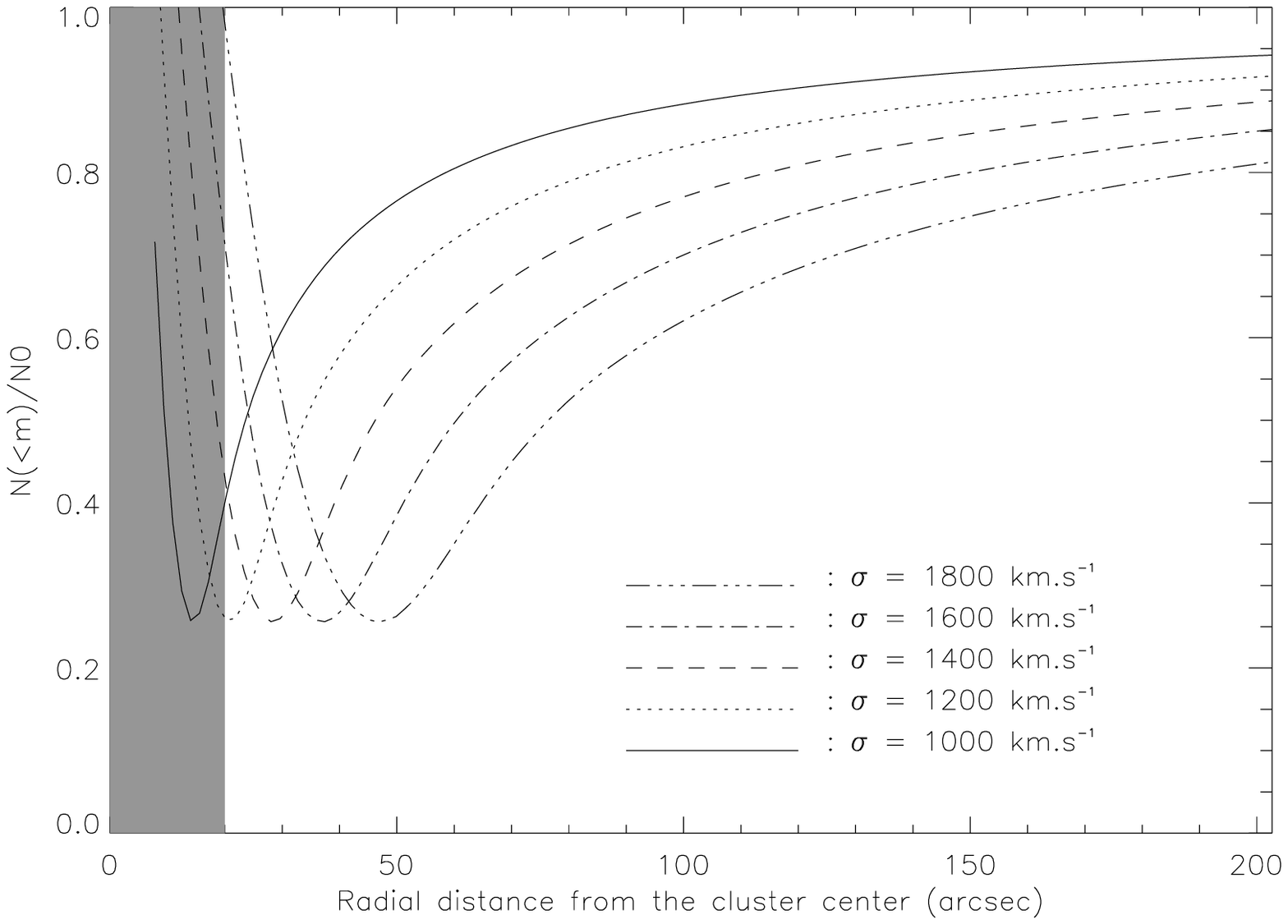}}}
\centerline{
\resizebox{\hsize}{!}{
\includegraphics{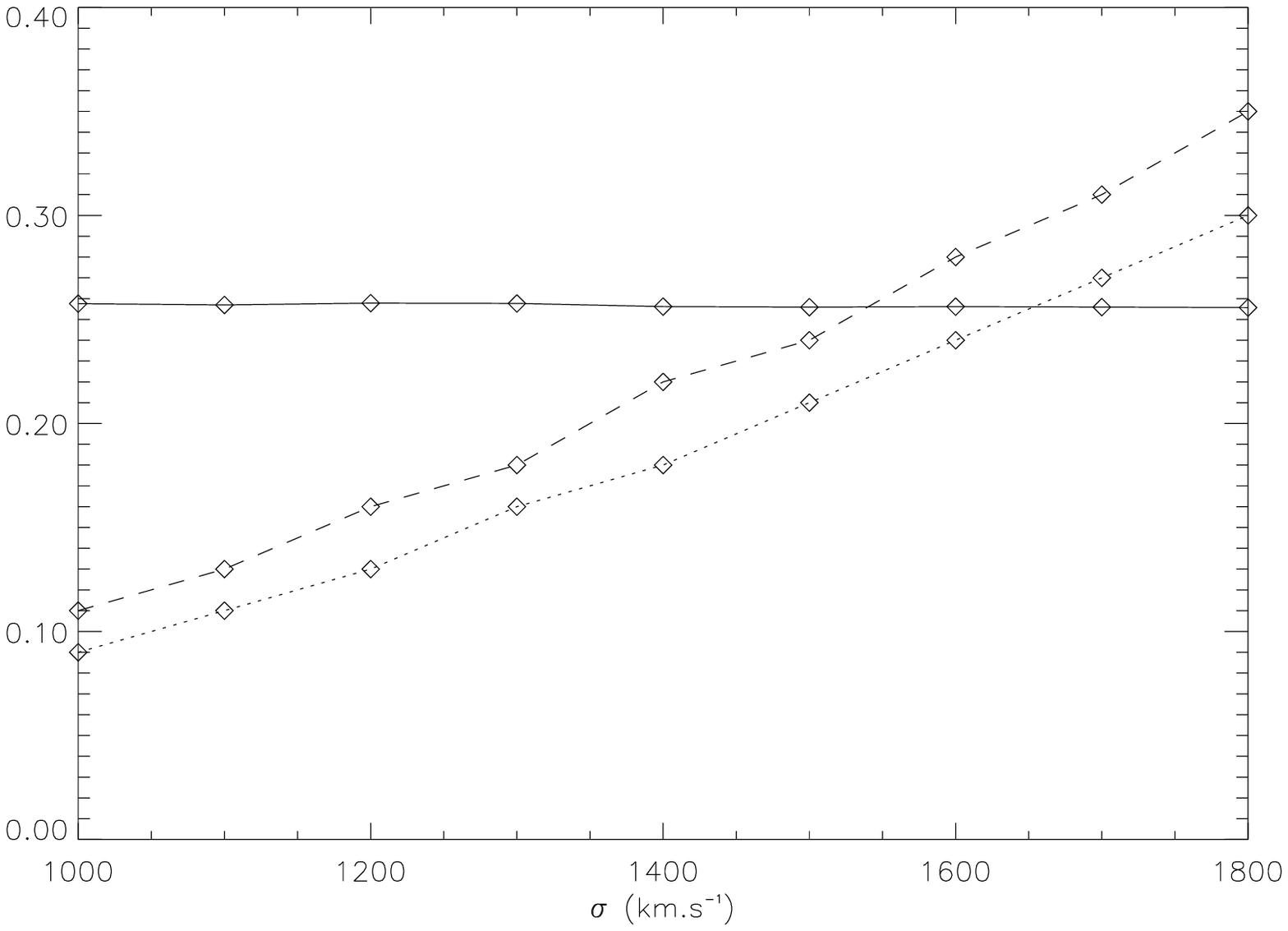}}}
\caption{{\bf Top: } Depletion curves obtained for different velocity
dispersions ranging from 1000 to 1800 km s$^{-1}$.  {\bf Bottom: }
Intensity of the minimum (solid line), position of the minimum in Mpc
(dotted line) and half width at half minimum in Mpc (dashed line) of
the depletion curves as a function of the velocity dispersion.}
\label{dep_sigma}
\end{figure}

We used a SIS model and varied the velocity dispersion $\sigma$ from
1000 to 1800 km s$^{-1}$ (Fig. \ref{dep_sigma}). As expected the
radius of the minimum increases with $\sigma$. Indeed for a SIS, the
Einstein radius scales exactly as $\sigma^2$, so the minimum of the
depletion curve, which roughly corresponds to the maximum Einstein
radius (at $z \sim 4$ or more), also scales as $\sigma^2$.

More surprisingly, the depth of the depletion at the position of
the minimum is roughly constant and, at first order, does not depend
on the velocity dispersion of the cluster or its total mass. Part of
this effect is due to the density of foreground galaxies, but part of
it is intrinsic to this SIS model, as other potential shapes do not
show this property (see below). On the contrary there is a clear
dependence on the half width at half minimum with $\sigma$ (Figure
\ref{dep_sigma}). 

\subsection{Influence of the core radius} 
\begin{figure}
\centerline{
\resizebox{\hsize}{!}{
\includegraphics{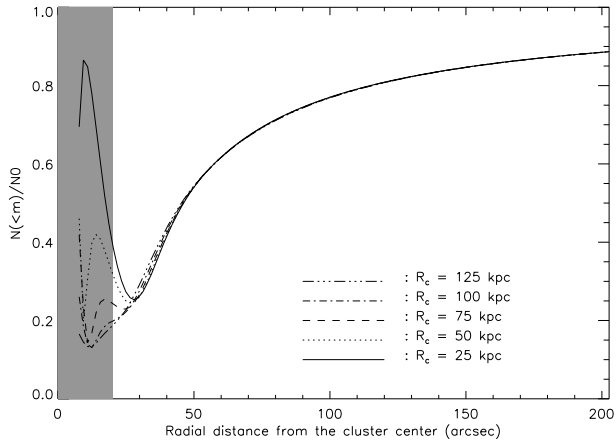}}}
\caption{Depletion curves obtained with $\sigma=1400$ km s$^{-1}$ for
different core radii ranging from 25 to 125 $h_{50}^{-1}$ kpc. The
value of the Einstein radius at the cluster redshift is $\sim$ 200
$h_{50}^{-1}$ kpc.}
\label{dep_rc}
\end{figure}

The introduction of a core radius in the model (Fig. 
\ref{dep_rc}) does not affect significantly the outer region of the
depletion area and has little effect on the position of the minimum.
As $R_C$ increases, the inner width of the curve is enlarged and its
slope is decreased. In fact, the study of this area for our purpose
has no interest, because the spatial resolution of galaxy number
counts variations is by far larger than this scale. In addition, in
rich clusters of galaxies, the central density of galaxies is large
enough so that empty areas are quite small between the envelopes of
large and bright galaxies. 

In practice, the only observable effect of a core radius is a
deformation of the inner depletion curve with a significant departure
from a symmetry with respect to the outer part. But we suspect that a
quantitative estimate of $R_C$ would be difficult to extract from the
signal. 

\subsection{Influence of the slope of the mass profile}
\begin{figure}
\centerline{
\resizebox{\hsize}{!}{
\includegraphics{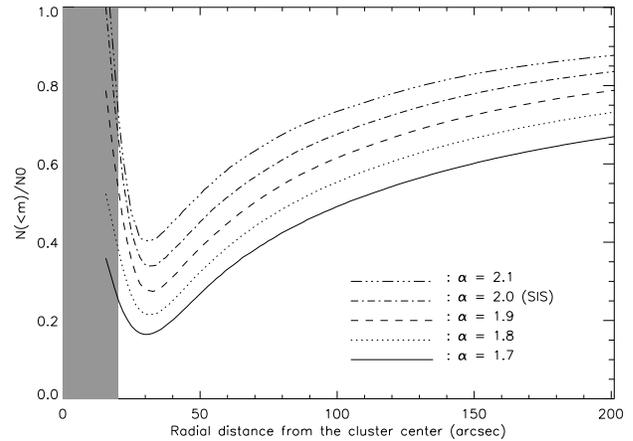}}}
\centerline{
\resizebox{\hsize}{!}{
\includegraphics{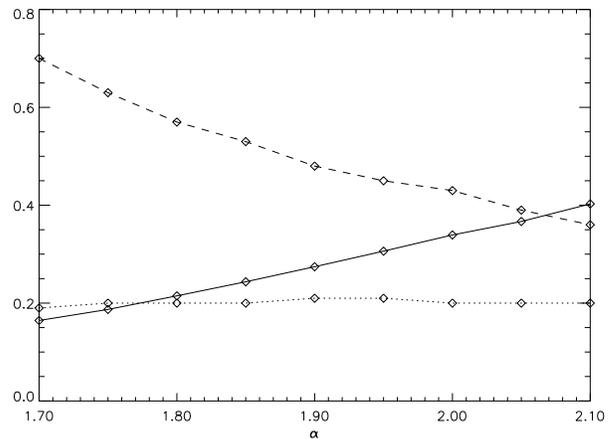}}}
\caption{{\bf Top: } Depletion curves obtained with $\rho_E = 2.25\times
10^{15}$ M$_{\sun}/$Mpc$^3$ and for different slopes $\alpha$ ranging
from 1.7 to 2.1.  {\bf Bottom: } Intensity of the minimum (solid
line), position of the minimum in Mpc (dotted line) and half width at
half minimum in Mpc (dashed line) of the depletion curves as a
function of the slope $\alpha$.}
\label{dep_alpha}
\end{figure}

The simulations are done with the power-law density profile,
characterised by its slope $\alpha$ (Fig. \ref{dep_alpha}).  The
minimum position of the depletion area does not depend significantly
of $\alpha$, because our scaling of the potential was fixed at the
Einstein radius, nearly independent of $\alpha$. On the contrary, the
two other typical features of the depletion area (half width at half
minimum and intensity of the minimum) strongly depend on this parameter. The
width of the curve represents roughly the mass dependence with
radius. In the case $\alpha = 1.7$ for example, the shallower slope of
the mass radial dependence creates a larger width of the depletion
curve.  The half width at half minimum of the curve decreases when
$\alpha$ increases. The reason is the same as with the SIS, that is to
say, when $\alpha$ increases, the mass is concentrated in the inner
part of the cluster and the depletion effect is less extended to the
outer regions.

\subsection{Influence of the ellipticity of the potential}
\label{ellipticity}
\begin{figure}
\centerline{
\resizebox{\hsize}{!}{
\includegraphics{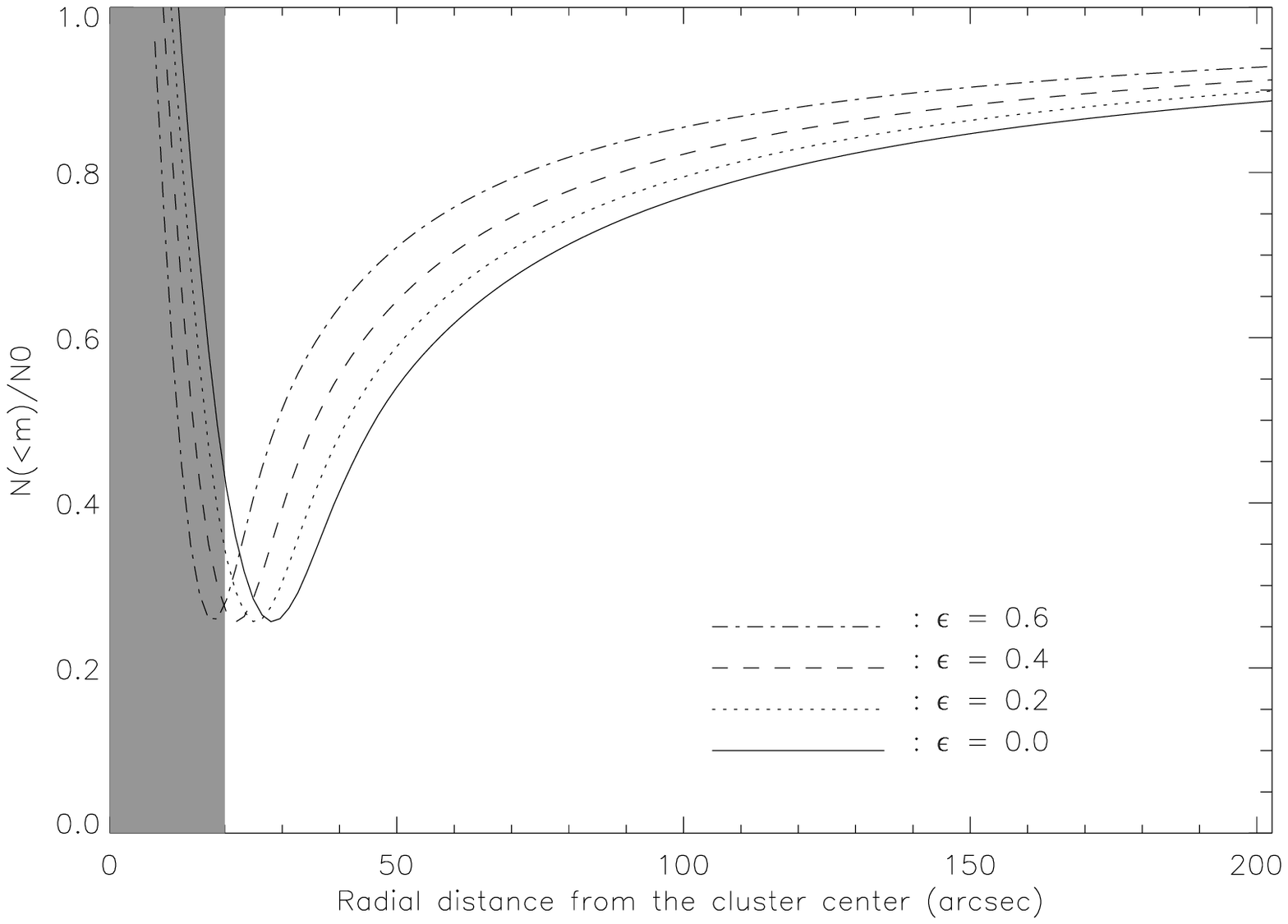}}}
\centerline{
\resizebox{\hsize}{!}{
\includegraphics{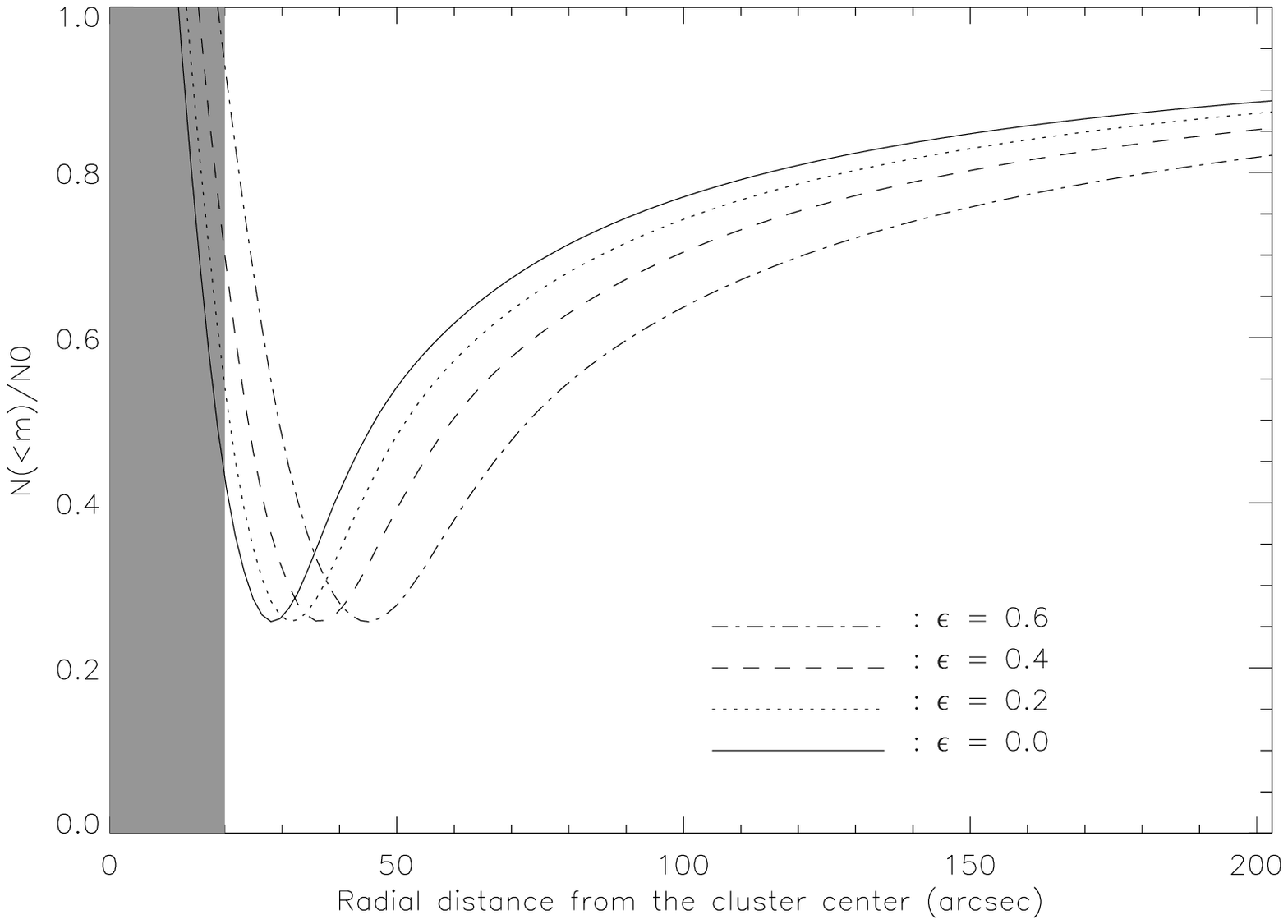}}}
\caption{{\bf Top: } Variation of the depletion curve with ellipticity 
along the minor axis. {\bf Bottom: } Idem along the major axis. The
streching of the curves with ellipticity is characterised by an
homothetic transformation.}
\label{dep_epsilon1}
\end{figure}

In these simulations, we use the singular isothermal ellipsoid
potential, with $\sigma=1400$ km s$^{-1}$, and we study the depletion
curves along the minor and major axis (Fig. \ref{dep_epsilon1}).  The
increase of $\epsilon$ does not affect the minimum value of the
depletion area but leads to an increase/decrease of the half width at
half minimum and of the minimum position along the major/minor axis as
an homothetic transformation. The relative positions of the two minima
of the depletion area along the main axis gives immediately the axis
ratio and thus the ellipticity of the potential.  An application of
this differential effect along the two main axis is presented in
Sect. \ref{applims1008}.

\subsection{Study of NFW profile} 
\begin{figure}
\centerline{
\resizebox{\hsize}{!}{
\includegraphics{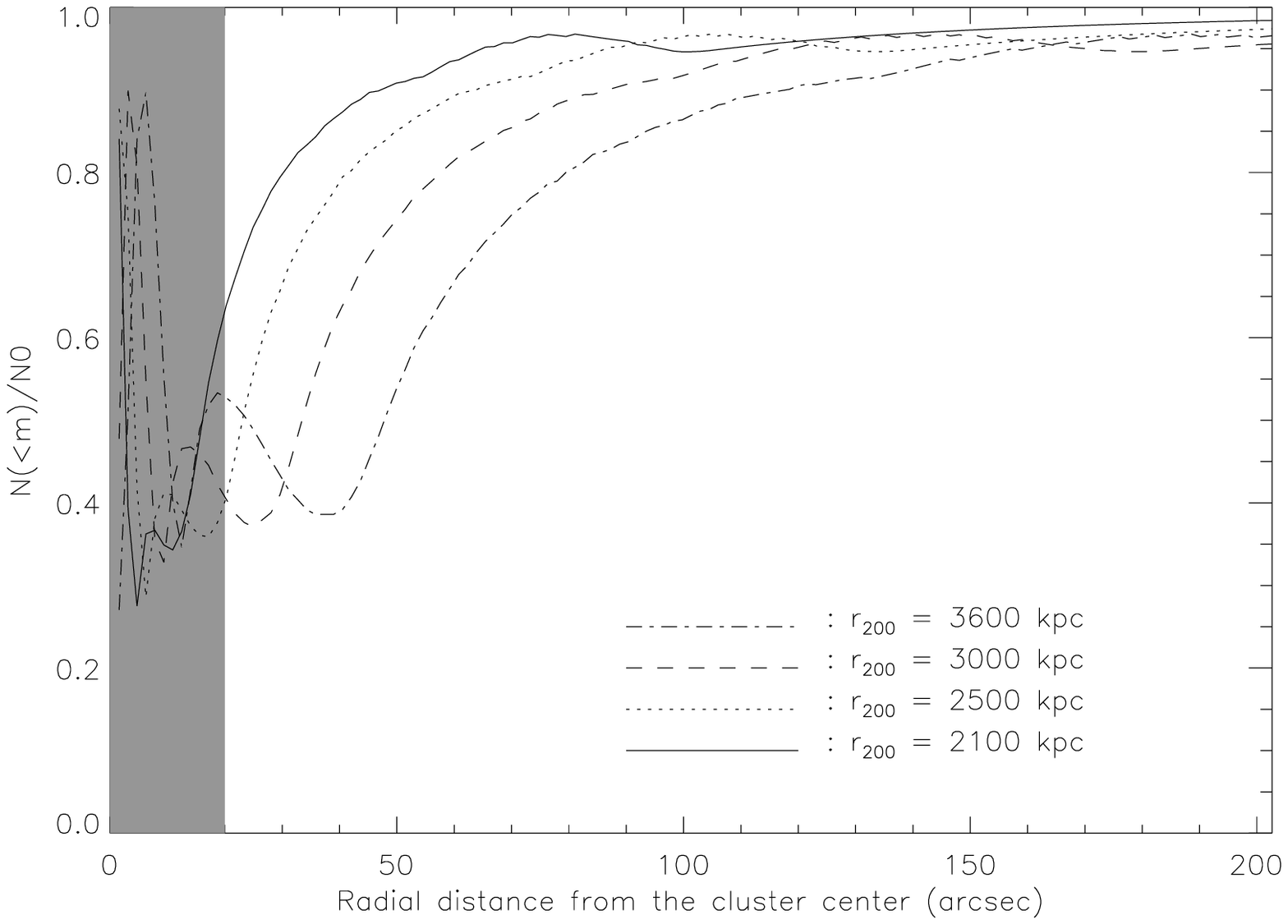}}}
\centerline{
\resizebox{\hsize}{!}{
\includegraphics{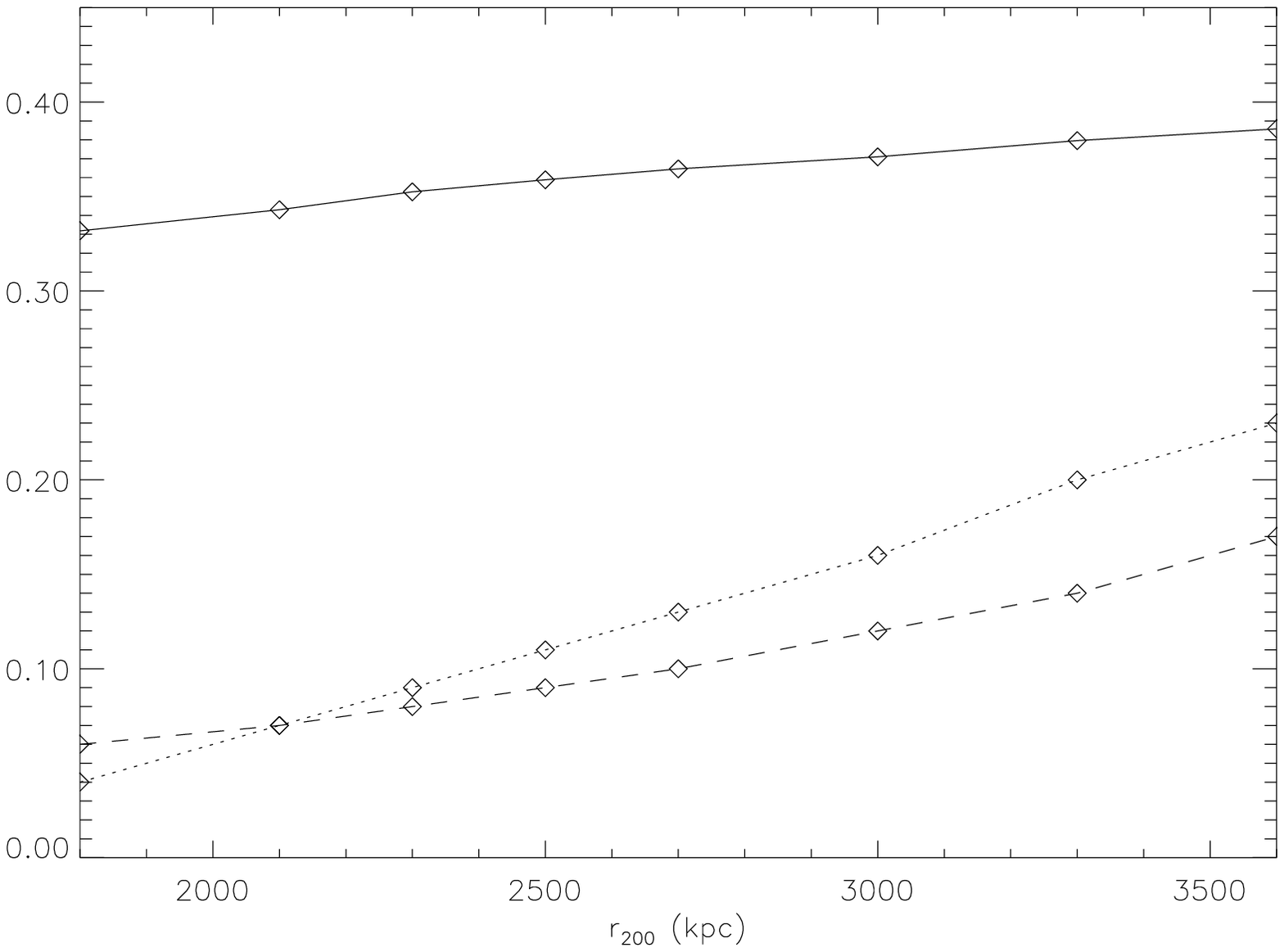}}}
\caption{{\bf Top: } Depletion curves obtained with $c=15$ and for different
virial radii. {\bf Bottom: } Intensity of the minimum (solid line),
position of the minimum in Mpc (dotted line) and half width at half
minimum in Mpc (dashed line) of the depletion curves as a function of
the virial radius.}
\label{depnfwc}
\end{figure}
\begin{figure}
\centerline{
\resizebox{\hsize}{!}{
\includegraphics{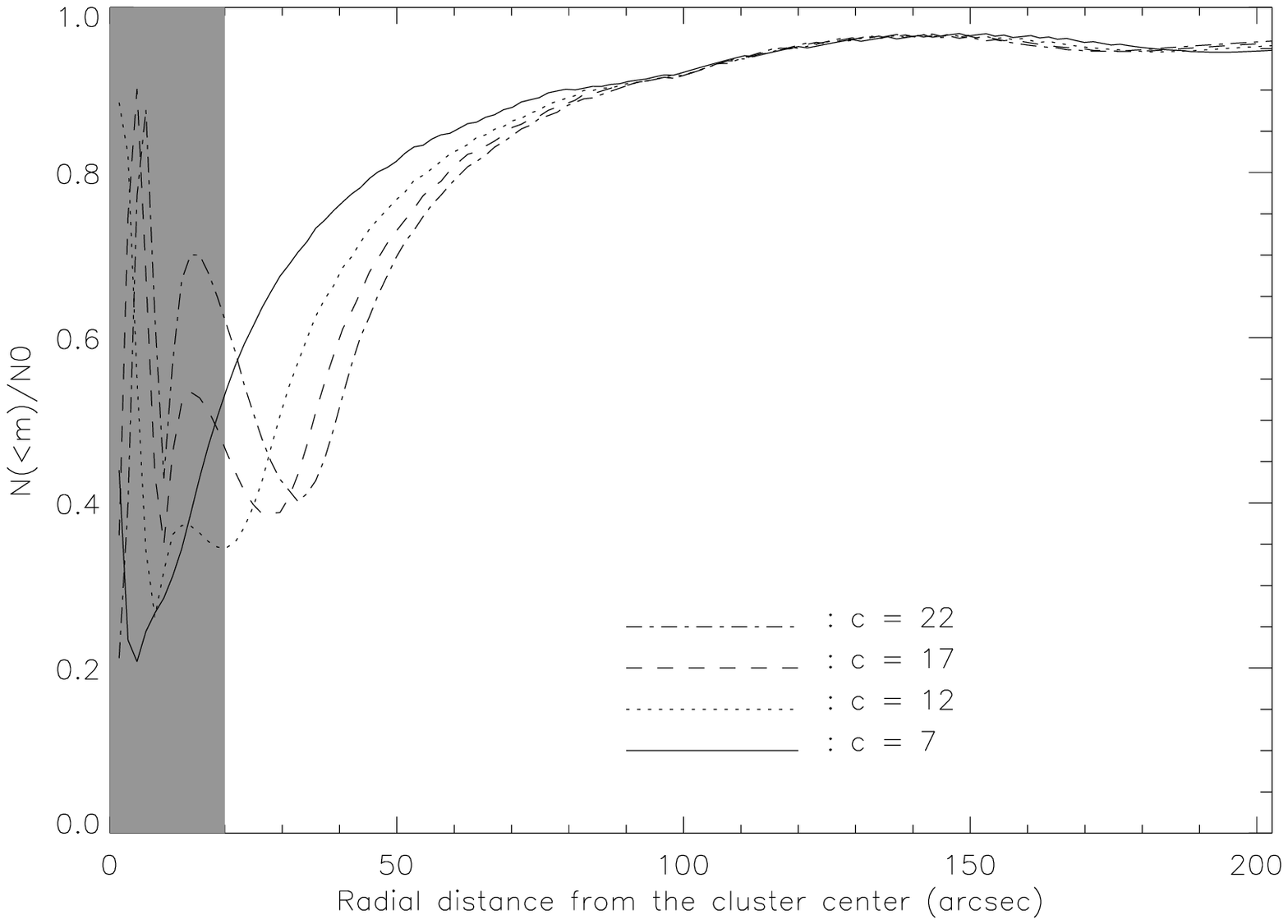}}}
\centerline{
\resizebox{\hsize}{!}{
\includegraphics{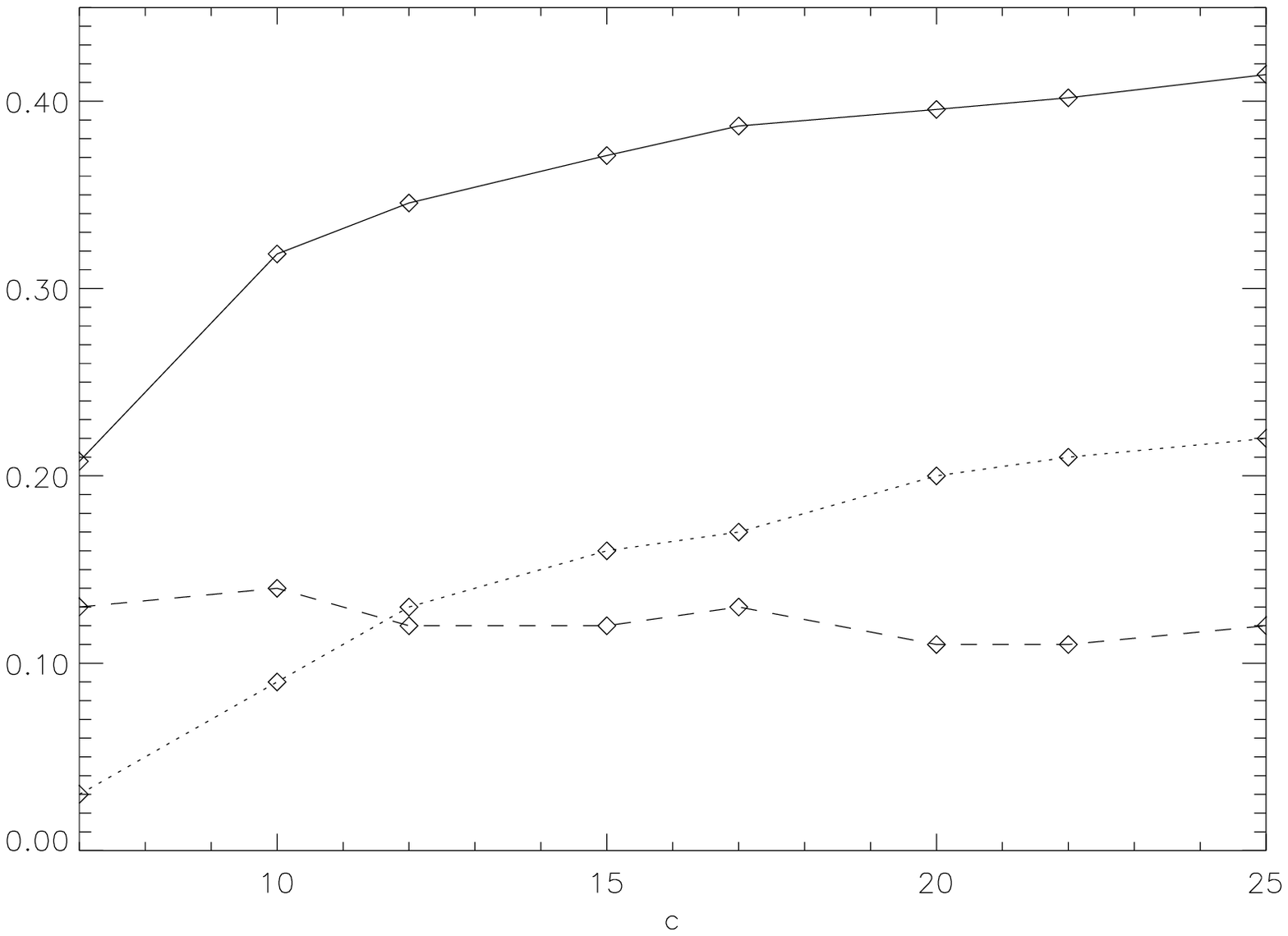}}}
\caption{{\bf Top: } Depletion curves obtained with $r_{200}=3000$ kpc and
for different concentration parameters. {\bf Bottom: } Intensity of
the minimum (solid
line), position of the minimum in Mpc (dotted line) and half width at half
minimum in Mpc (dashed line) of the depletion curves as a function of
the concentration parameter.}
\label{depnfwr}
\end{figure}
With the NFW model we varied separately the virial radius and the
concentration parameter. We have chosen the values for these two
parameters in the range of those found by Navarro et al. \cite*{navarro}. For the
first set of simulations, we adopted $c=15$ and varied $r_{200}$ from
1800 to 3600 kpc (Fig. \ref{depnfwc}). An increase of the virial
radius (more massive cluster) leads to an increase of the three
characteristic features of the depletion area and to the appearance of
a bump in the central region which grows with $r_{200}$. The increase
of the half width at half minimum of the depletion curves with
$r_{200}$ can be explained by the fact that when $r_{200}$ increases
there is more mass in the outer regions.  Consequently, the depletion
area is more extended to the outer part of the cluster and reaches its
asymptotic value less rapidly.  The second set of simulations was 
done with $r_{200}=3000$ kpc and we varied $c$ from 7 to 25
(Fig. \ref{depnfwr}). The variation of $c$ affects only the inner part
of the depletion curves. An increase of the concentration parameter
(cluster with smaller characteristic radius) leads to an increase of
the intensity and position of the minimum with a steeper slope for the small
values of $c$. Contrary to $r_{200}$, the variation of $c$ does not
affect significantly the half width at half minimum of the depletion
area.


\section{Influence of background galaxies on depletion curves}
In this section, we use a unique model for the lens, i.e. a SIS
profile with a velocity dispersion of 1500 km s$^{-1}$ for a cluster
located at $z_L = 0.4$, to avoid effects due to the lens.

\subsection{Effect of the lens on the magnitude-redshift distribution
of background populations} 
\begin{figure*}
\centerline{
\resizebox{\hsize}{!}{
\includegraphics{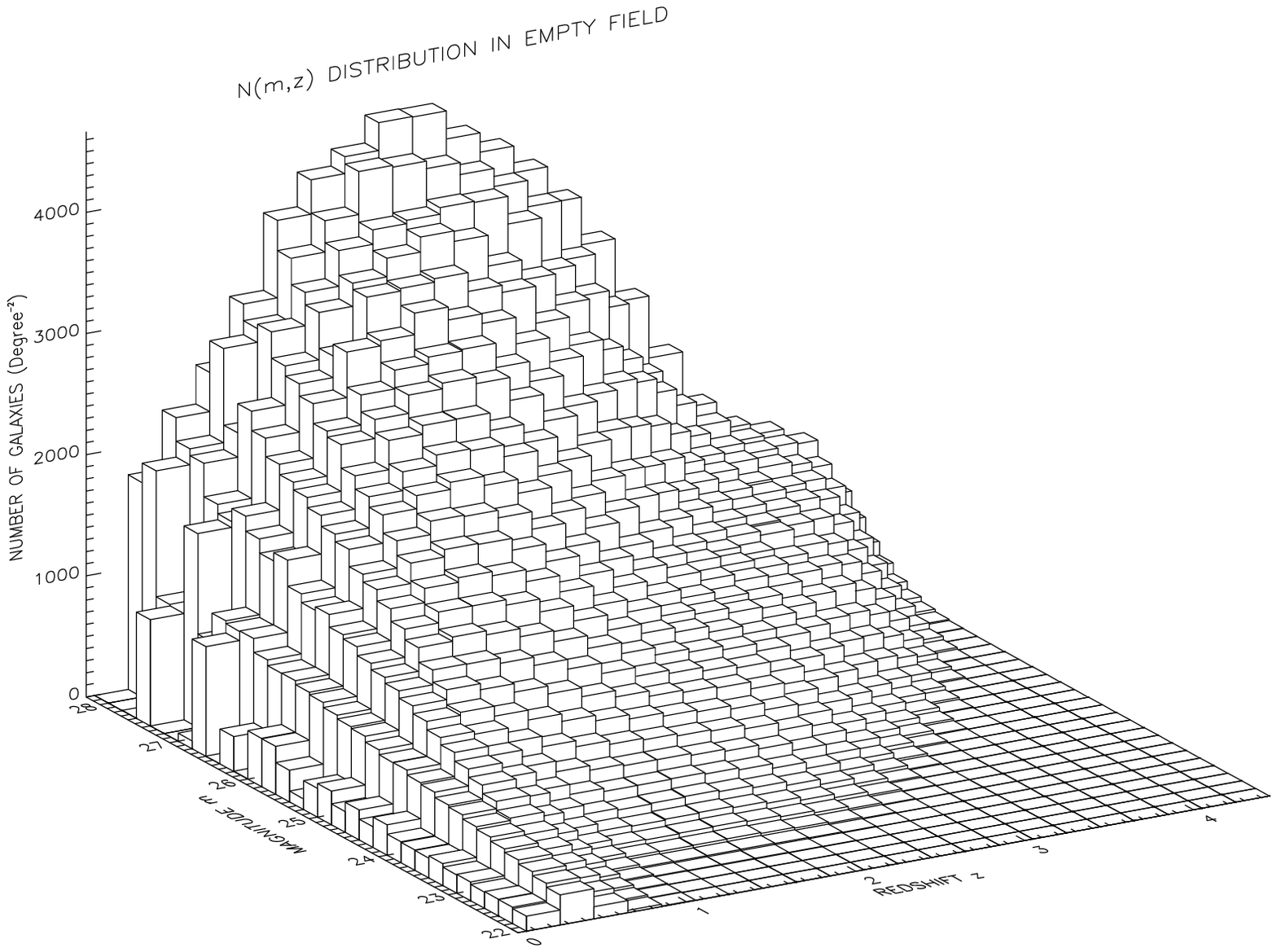}
\includegraphics{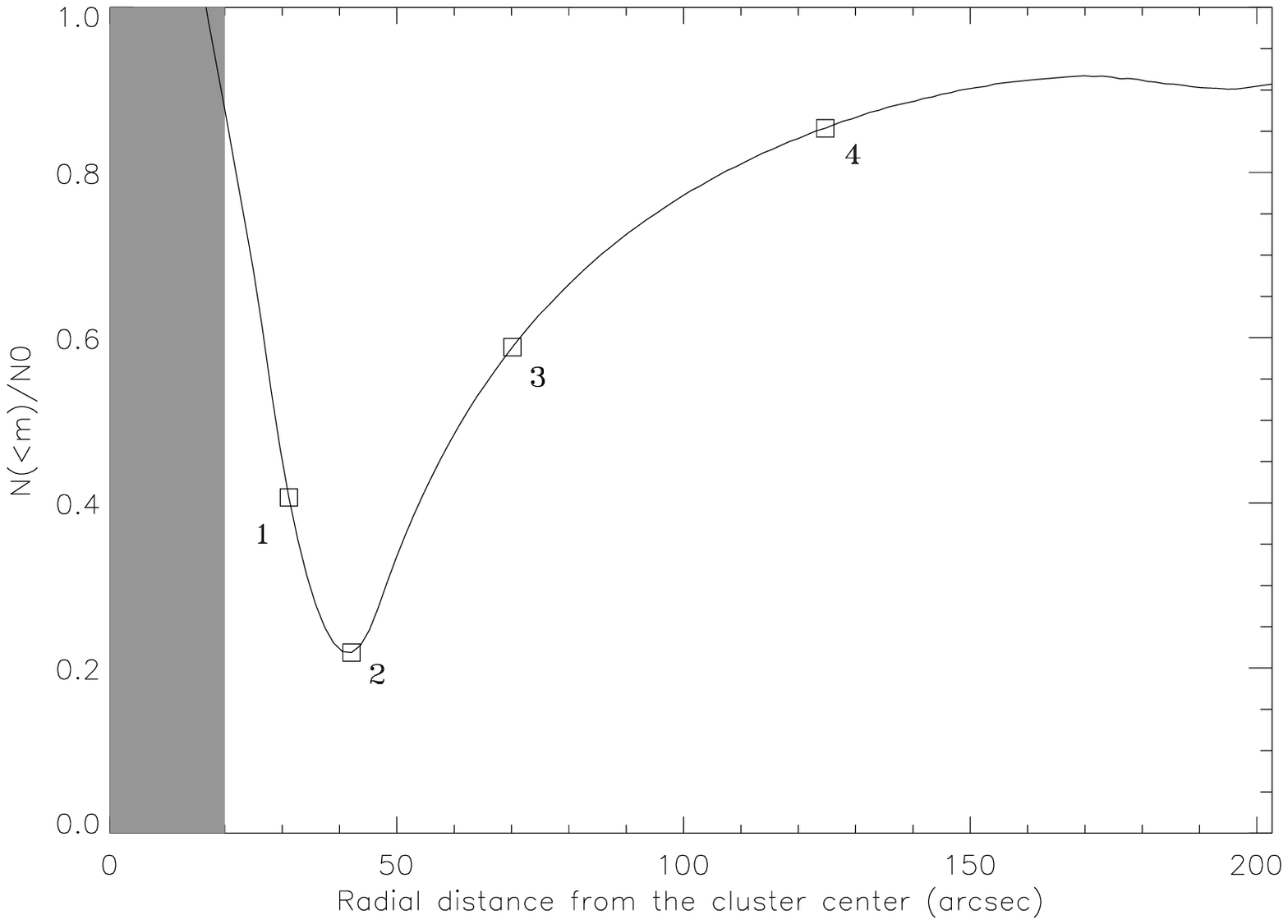}}}
\centerline{
\resizebox{\hsize}{!}{
\includegraphics{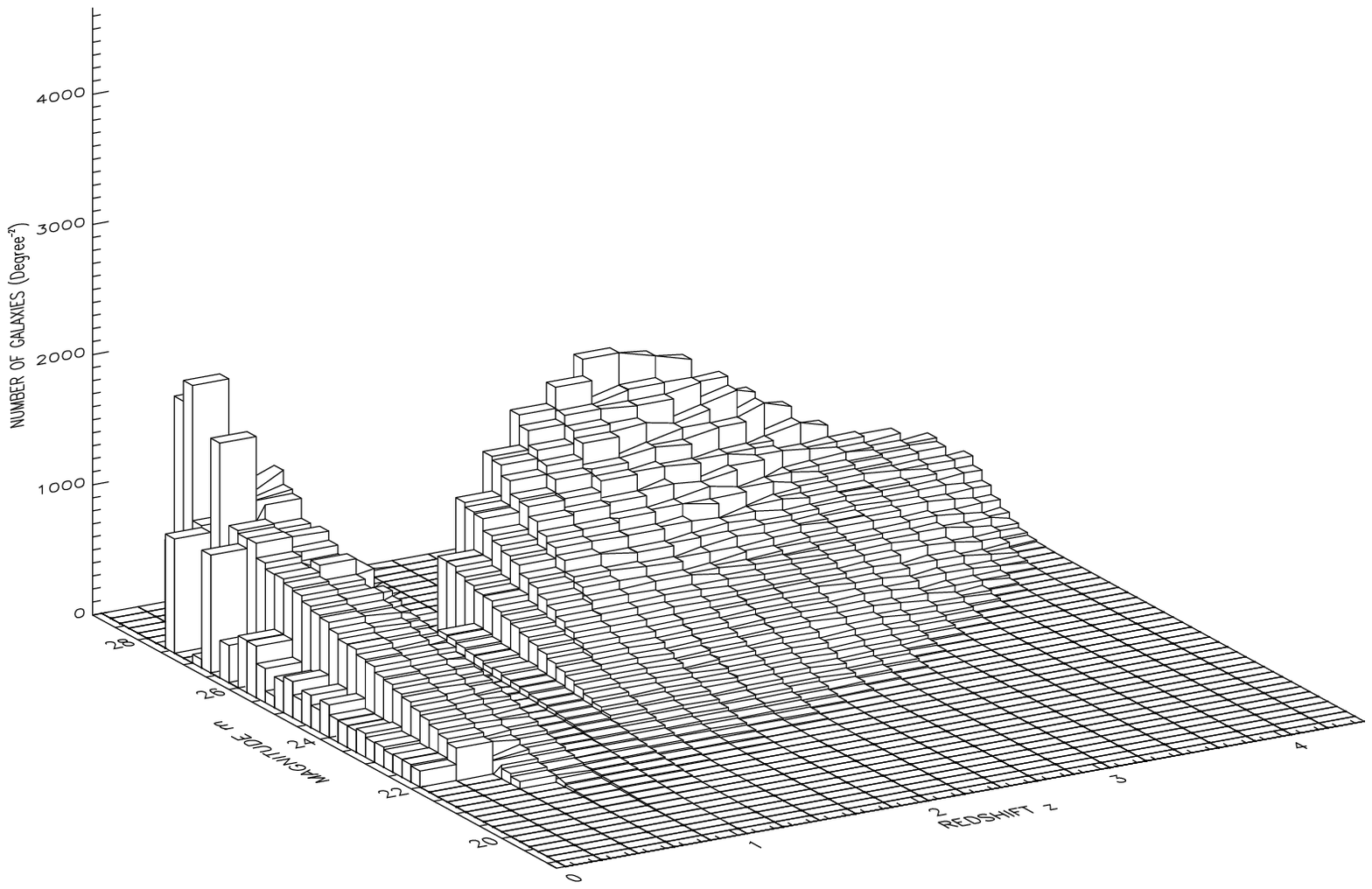}
\includegraphics{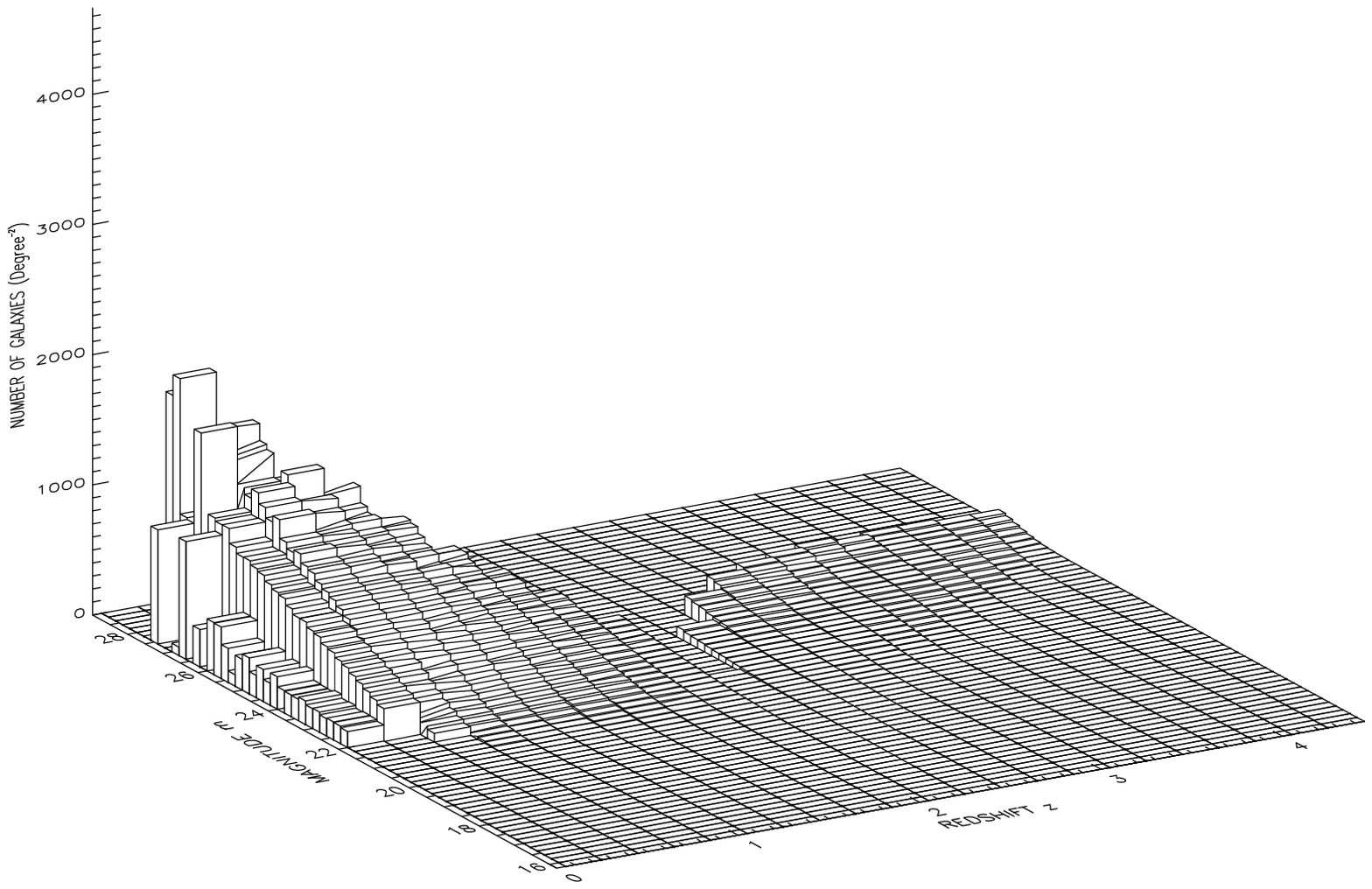}}}
\centerline{
\resizebox{\hsize}{!}{
\includegraphics{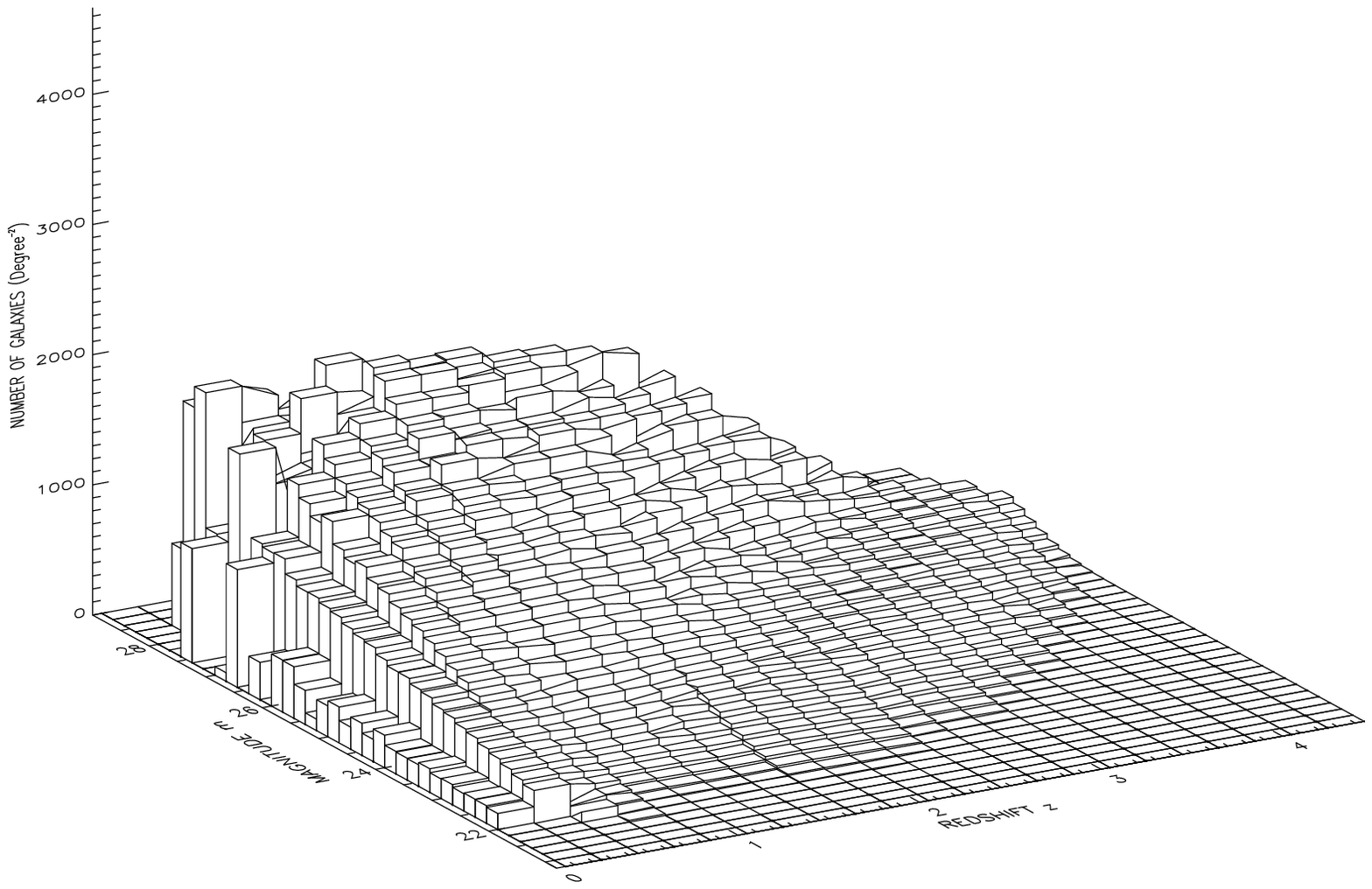}
\includegraphics{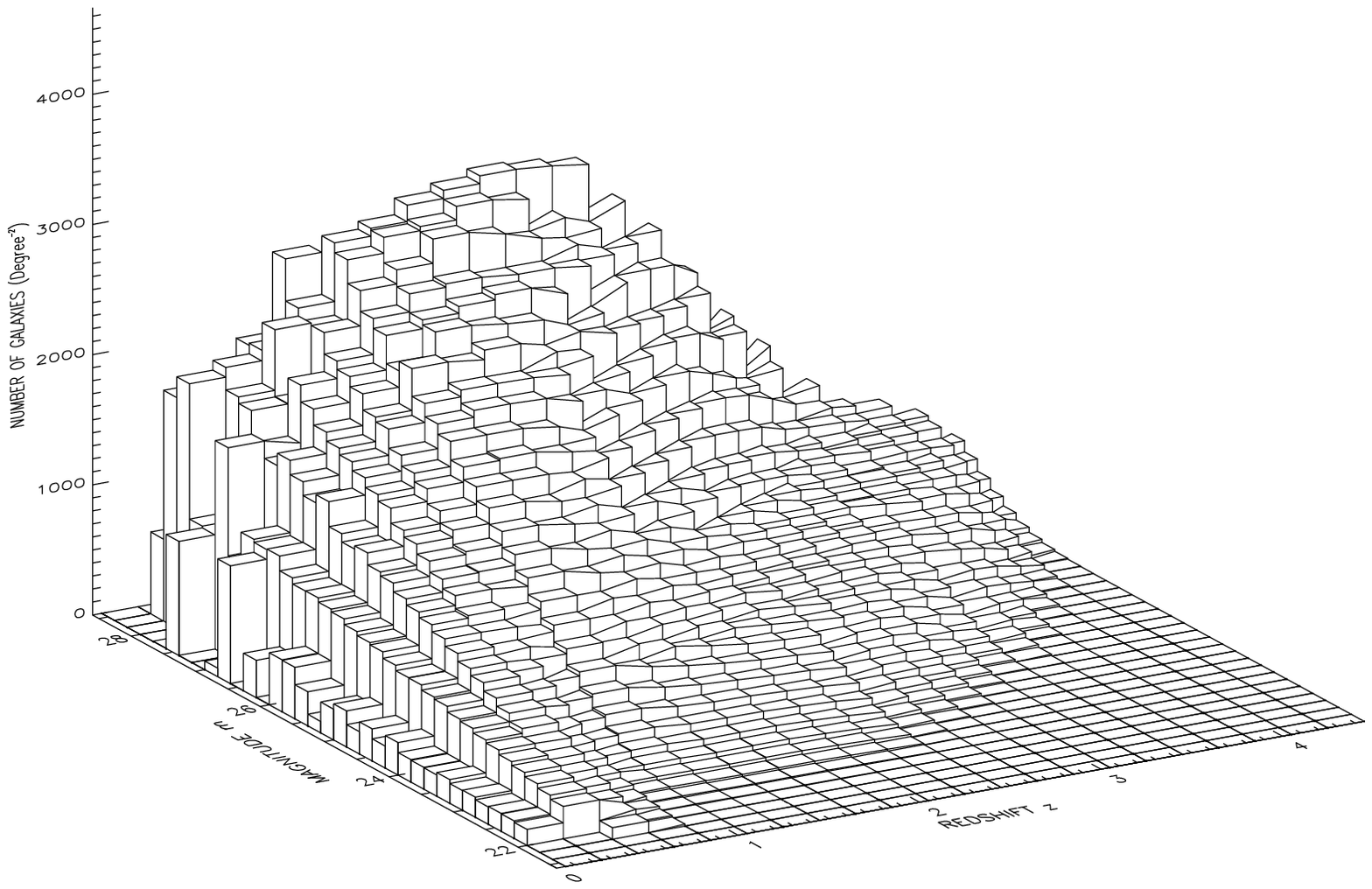}}}
\caption{{\bf Upper-left: } Magnitude-redshift distribution of the
galaxies in empty field computed from the model of B\'ezecourt et al. (1998), in   
the B-band; {\bf Upper-right: } Depletion
curve for $m_{Blim} =28$. The open squares show the points where the
following $N_{lens}(m,z,r)$ distributions are computed.  {\bf
Middle-left: } $r=31\arcsec$ (1); {\bf Middle-right: } $r=42\arcsec$
(2); {\bf Lower-left: } $r=70\arcsec$ (3); {\bf Lower-right: }
$r=125\arcsec$ (4).}
\label{NmzR}
\end{figure*}

Fig. \ref{NmzR} shows the distortion of the magnitude-redshift
distribution in the B band along the curve.  The results are
qualitatively the same in the others bands. Near the cluster center,
where the effects of the gravitational magnification are stronger,
objects located just behind the lens show a gain in magnitude larger
than 1. In the minimum area of the curve, the competition between
gravitational magnification and dilatation is more favorable to high
$z$ objects ($z>2.4$). In the outer part of the cluster, the effects
of the gravitational magnification decrease and one finds again the
distribution in empty field (asymptotic limit of the depletion
curves).

\subsection{Differential effects in several filters} 
\begin{figure*}
\centerline{
\resizebox{\hsize}{!}{
\includegraphics{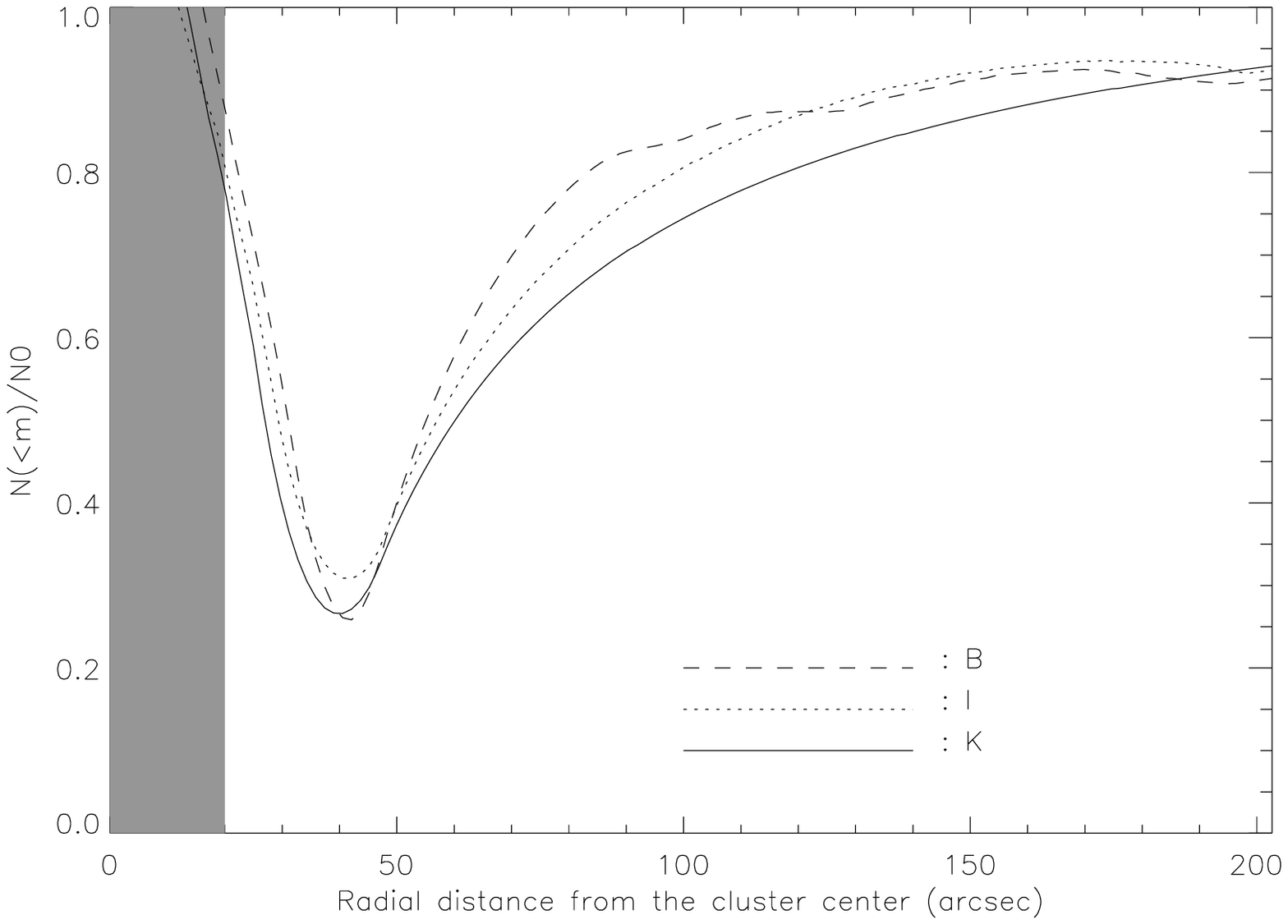}
\includegraphics{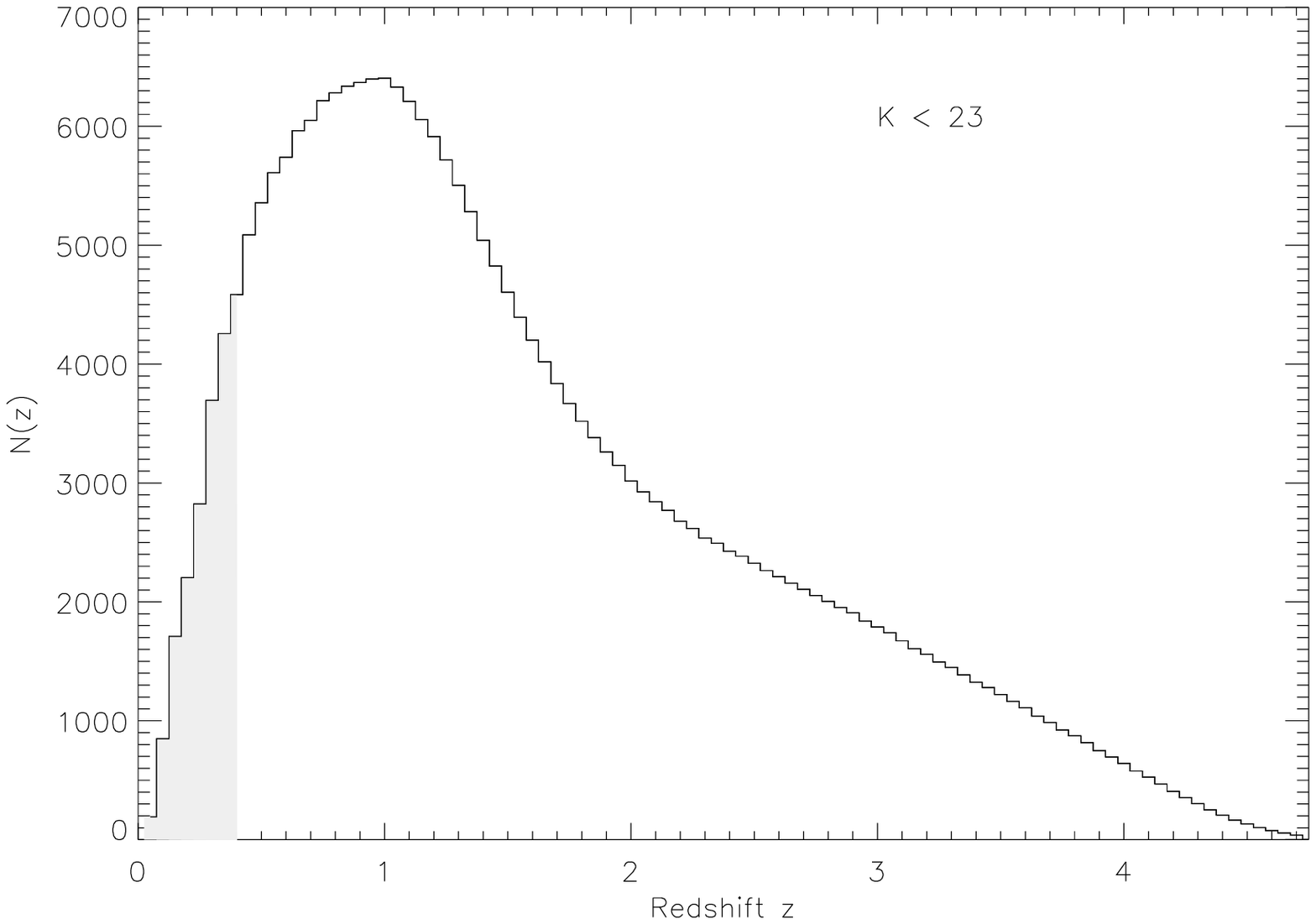}}}
\vspace*{0.5cm}
\centerline{
\resizebox{\hsize}{!}{
\includegraphics{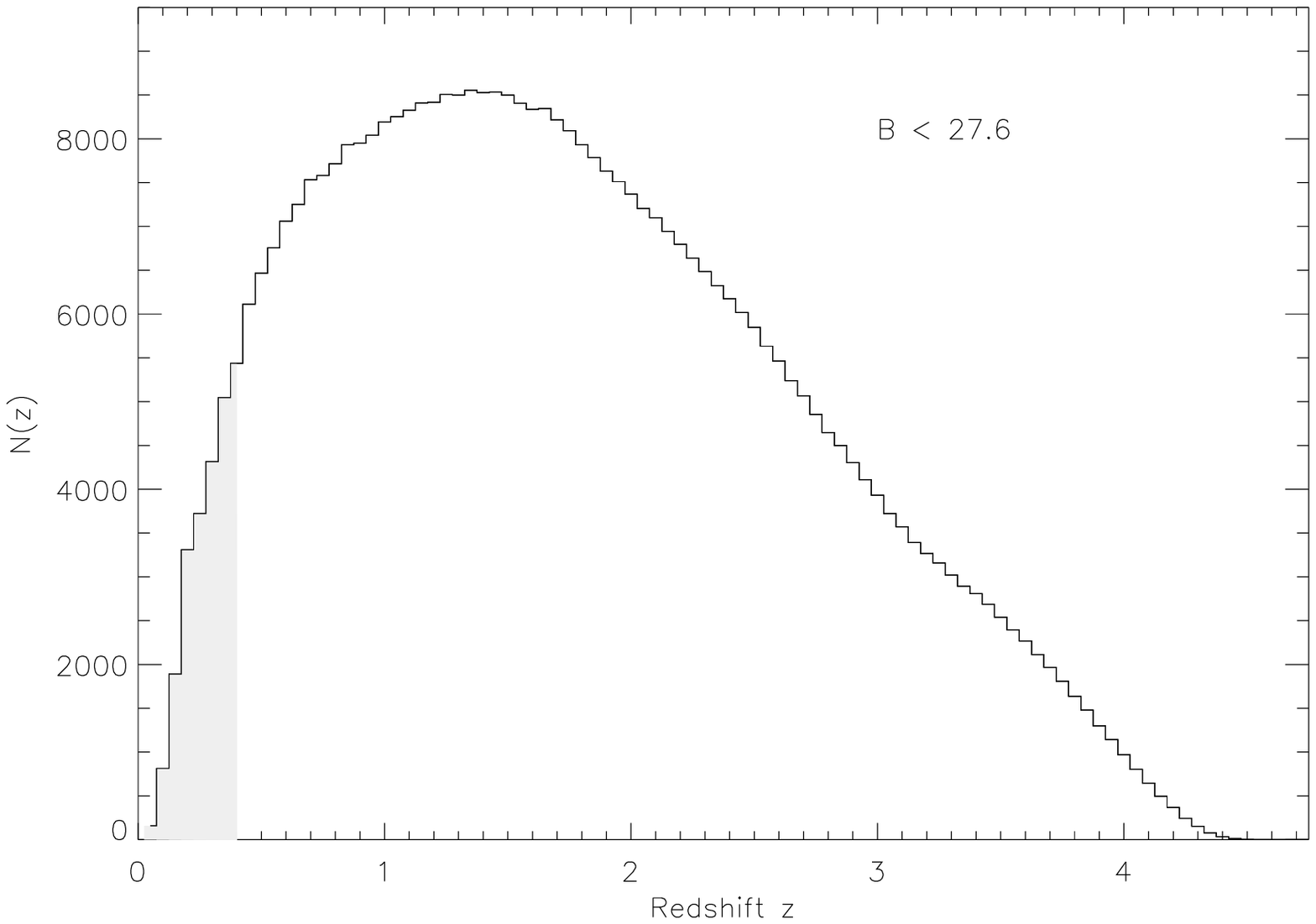}
\includegraphics{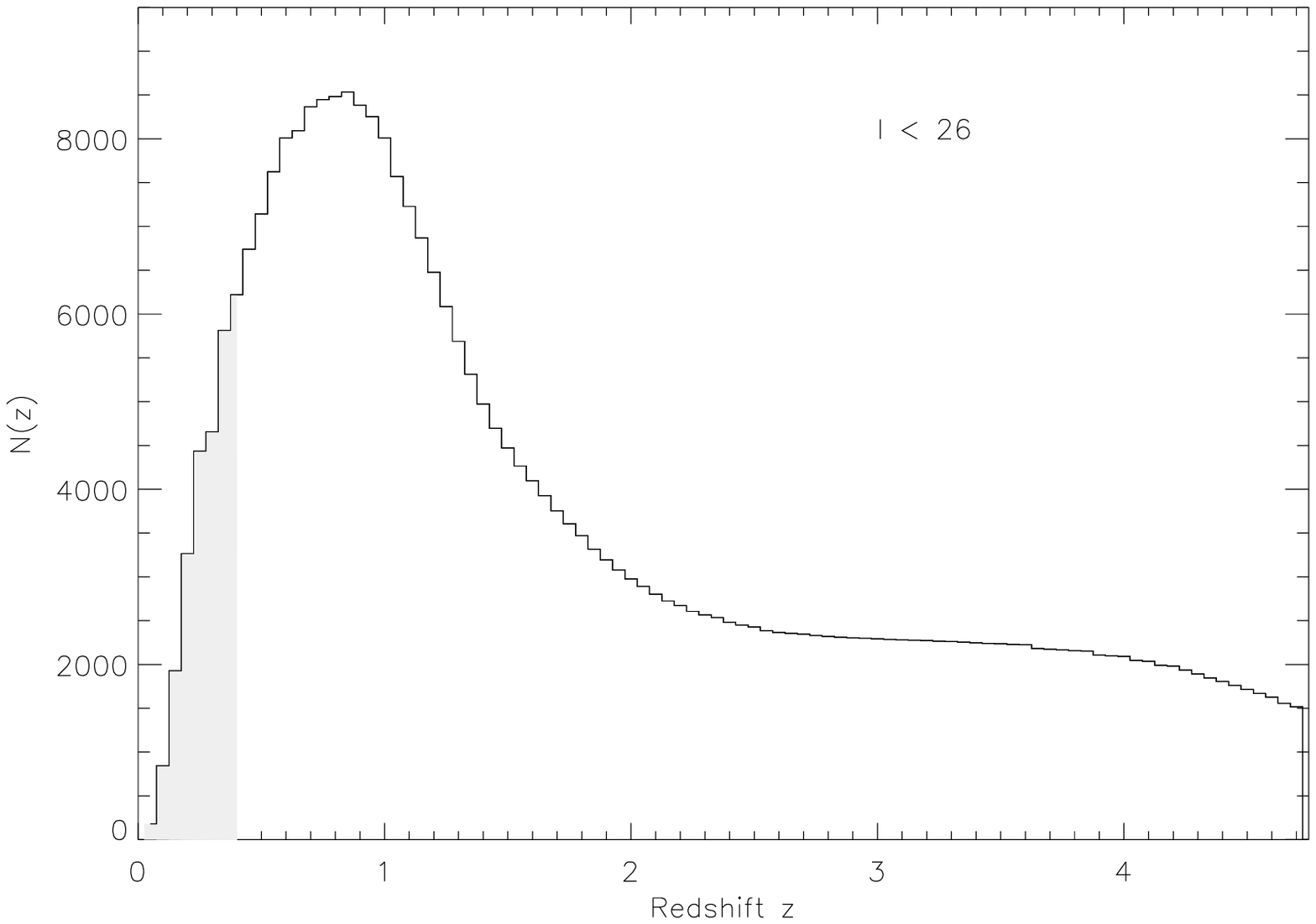}}}
\caption{Depletion curves obtained in different filters for a SIS
profile with $\sigma=1500$ km s$^{-1}$ (top left) and corresponding
field redshift distributions for each filter and magnitude limit. The
shaded area on the redshift distributions indicates the foreground
contaminating population.}
\label{effet_filtre}
\end{figure*}
We first studied the evolution of the depletion area with the redshift
distribution of the background population by considering several
filters. We tried to determine some characteristics of the redshift
distribution of background population which could be connected to
typical features of the depletion area (Fig. 
\ref{effet_filtre}).  The magnitude limits of the simulations correspond
to a typical observation time of about 2 hours on an 8-meter telescope
with good imaging facilities. The contamination by foreground objects
$( z < 0.4 )$ in the 3 filters is weak ($\simeq 8
\%$ in the I and K bands and $\simeq 6 \%$ in
the B band for which the counts are deeper).\\ The evolution of the
position of the minimum of the depletion area reflects the evolution of
the mean redshift of the different populations ($<z>\simeq 1.57$
for the K band, $<z>\simeq 1.76$ for the B band and $<z>\simeq
1.79$ for the I band), but with a weak effect. The variation of the
intensity of the minimum of the depletion area is connected for one part to
the fraction of foreground objects and for the other part to the
slope $\gamma$ of the luminosity function.  As this slope decreases
from I band to B band $(\gamma_I=0.23, \gamma_K=0.22,
\gamma_B=0.14)$ with the magnitude thresholds we have chosen, the
intensity of the minimum of the depletion area decreases in the same way from I
band ($\simeq 0.31$) to B band ($\simeq 0.26$). Both features of the
depletion curves are in fact weakly dependent on the choice of the
filter and may be difficult to distinguish with real data, at least
while the variation of the redshift distribution with wavelength is
rather smooth. \\ On the contrary, the half width at half minimum of
the depletion curve is affected by the redshift distribution of
background populations in a more sensitive way.  The increase of the
half width at half minimum from B band to K band is anti-correlated
with the fraction of objects at high redshift ($z>2$) for these
populations which also decreases in the same way from B band ($\simeq
37 \%$) to K band ($\simeq 29 \%$) and seems to reflect essentially
the bulk of galaxies at redshift around 1.


\section{Depletion effect in the cluster MS1008--1224}
\label{applims1008}

We have applied our simulations to the cluster MS1008--1224
($z=0.3062$, Lewis et al. \cite*{lewis}) which is one cluster of the
\textit{Einstein Observatory} Extended Medium Sensitivity Survey
\cite{gioia}.  It is a very rich galaxy cluster, slightly extended in
X-rays, and it is also part of the Canadian Network for 
Observational Cosmology Survey \cite{carlberg}. Its
galaxy distribution is quite circular, surrounding a
North-South elongated core.  There is a secondary clump of galaxies to
the North. The X-ray luminosity is L$_X$(0.3-3.5 keV)$ = 4.5 \times
10^{44}$ erg.s$^{-1}$ \cite{gioia}, the X-ray temperature inferred
from ASCA observations is T$_X=7.3$ keV \cite{mushotzky} and the
radio flux at 6 cm is lower than 0.8 mJy \cite{gioia}. Some
gravitationally lensed arcs to the North and to the East of the field
have been reported by Le F\`evre et al. \cite*{lefevre} as well as by 
Athreya et al. \cite*{athreya}.
\begin{figure}
\centerline{
\resizebox{\hsize}{!}{
\includegraphics{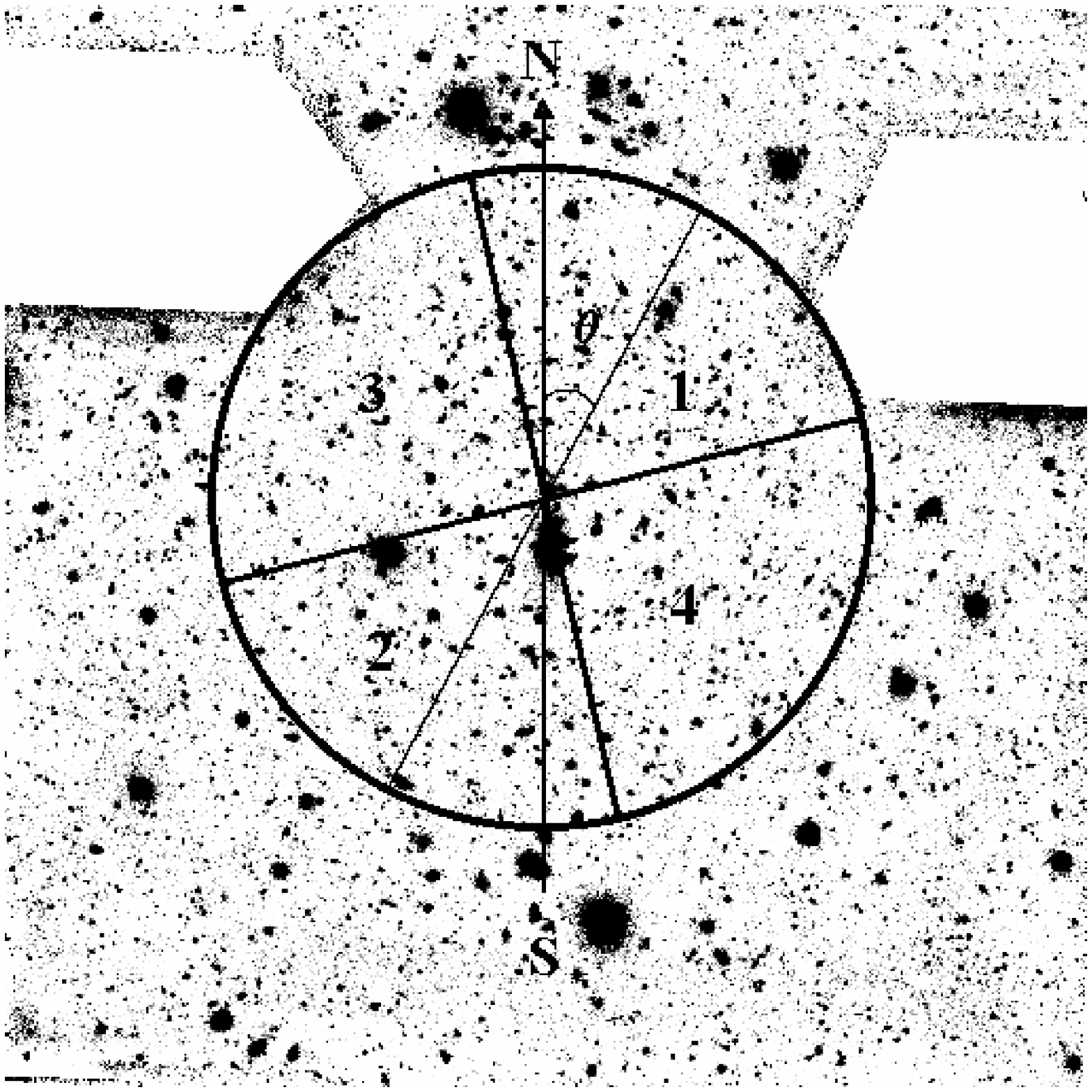}}}
\caption{VLT FORS1 image of MS1008--1224 in the B band. North is up,
East is left. The field-of-view is $6.8\arcmin \times 6.8\arcmin
$. Note the two occulting masks hiding the brightest stars in the
field.}
\label{ms1008B}
\end{figure}

\begin{figure}
\centerline{
\resizebox{\hsize}{!}{
\includegraphics{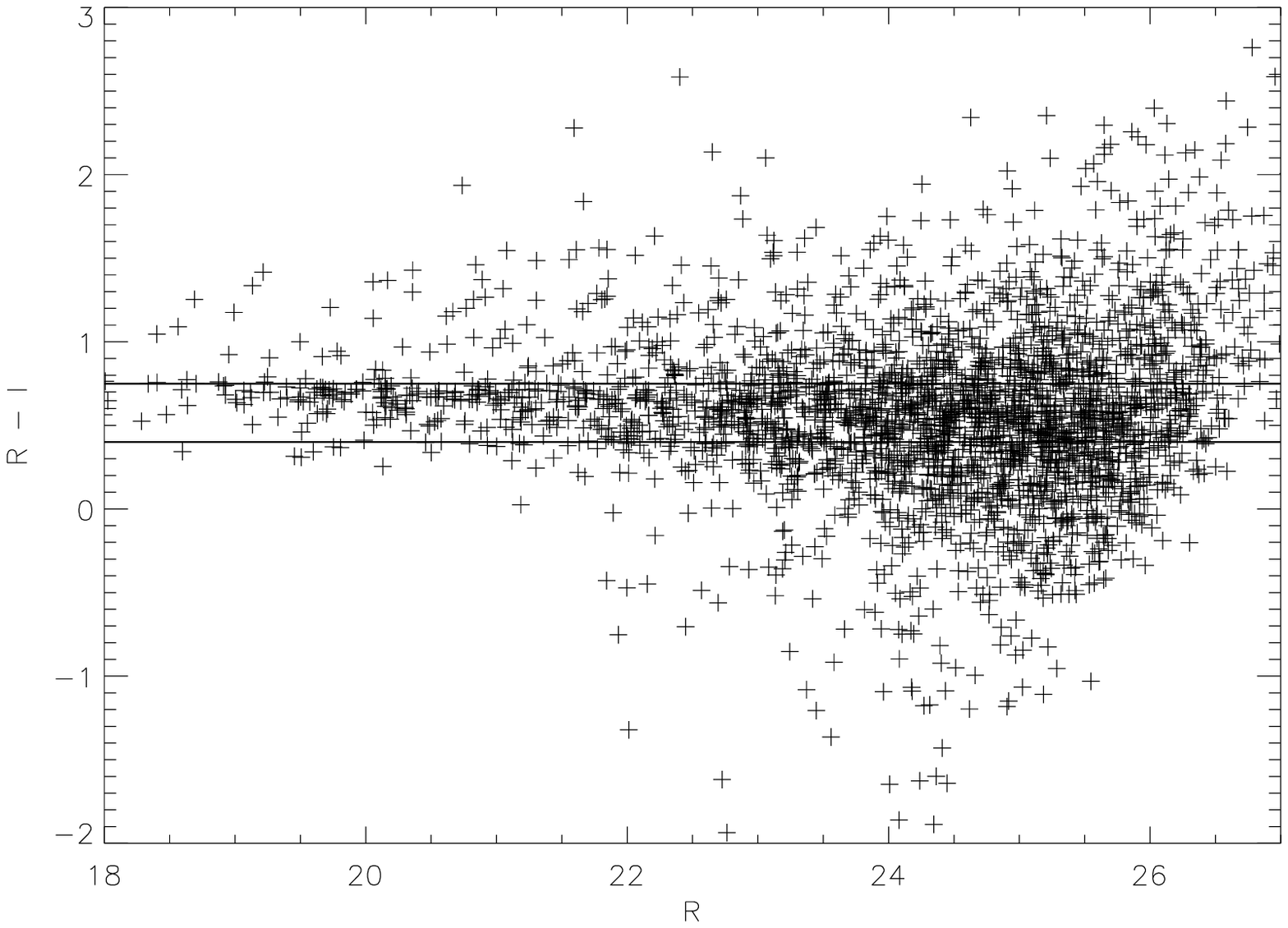}}}
\caption{Color-magnitude diagram R--I versus R of MS1008--1224. The two
horizontal lines delimit the sequence of the cluster elliptical
galaxies, which has been removed from our catalogue.}
\label{catalogue}
\end{figure}

\subsection{The observations} 
Data were obtained during very deep observations with FORS and ISAAC
during the science verification phase of the VLT--ANTU (UT1) at Cerro
Paranal ({\tt http://www.eso.org/science/ut1sv}).  Multicolor photometry
was obtained in the B, V, R and I bands with FORS1 (field-of-view $6.8'
\times 6.8'$) and in the J and Ks bands with ISAAC (field-of-view $2.5'
\times 2.5'$) with sub-arcsecond seeing in all cases. The total exposure
times are respectively 2880 seconds in J, 3600 seconds in Ks, 4050
seconds in I, 4950 seconds in B, 5400 seconds in V and R.
We used the SExtractor software \cite{bertin} to construct our
photometric catalogue.  The completeness magnitudes are
respectively $\mathrm{J}=24$, $\mathrm{Ks}=22$,
$\mathrm{I}=25.5$, $\mathrm{V}=26.5$, $\mathrm{B}=26.5$ and
$\mathrm{R}=26$. Our values are identical to those of Athreya et al. 
\cite*{athreya} concerning the FORS observations but they are one
magnitude deeper concerning the ISAAC ones. In practice, we used only
the FORS data which extend to a larger distance and are more suited to
our study. 

In order to remove part of the contamination by cluster members, we
identified the elliptical galaxies by their position on the
color--magnitude diagram (Fig. \ref{catalogue}). So background
sources were selected by removing this sequence, essentially valid at
relatively bright magnitudes. This statistical correction is not fully
reliable as it does not eliminate bluer cluster members or foreground
sources. But it partly removes one cause of contamination, especially
significant close to the cluster center.  Our counts may also present
an over-density of objects in the inner part of the cluster
($r<20\arcsec$) due to the non-correction of the surface of cluster
elliptical galaxies we have removed, effect which is more sensitive in
the inner part of the cluster where these galaxies are dominant. But
as it is not in the most interesting region of the depletion area, we
did not try to improve the measures there.

In addition, due to the presence of two 11 magnitude stars in the
northern part of the FORS field, two occulting masks were put to avoid
excessive bleeding and scattered light. We took into account this
partial occultation of the observed surface for the radial counts
above a distance of 130\arcsec\ (see Fig. \ref{ms1008B}).  This may
possibly induce some additional errors in the last two points of the
curves, which are probably underestimated, because of the difficulty
to estimate the surface of the masks and some edge effects at the
limit of the field.

\subsection{Depletion curves and mass density profile}
\begin{figure*}
\centerline{
\resizebox{\hsize}{!}{
\includegraphics{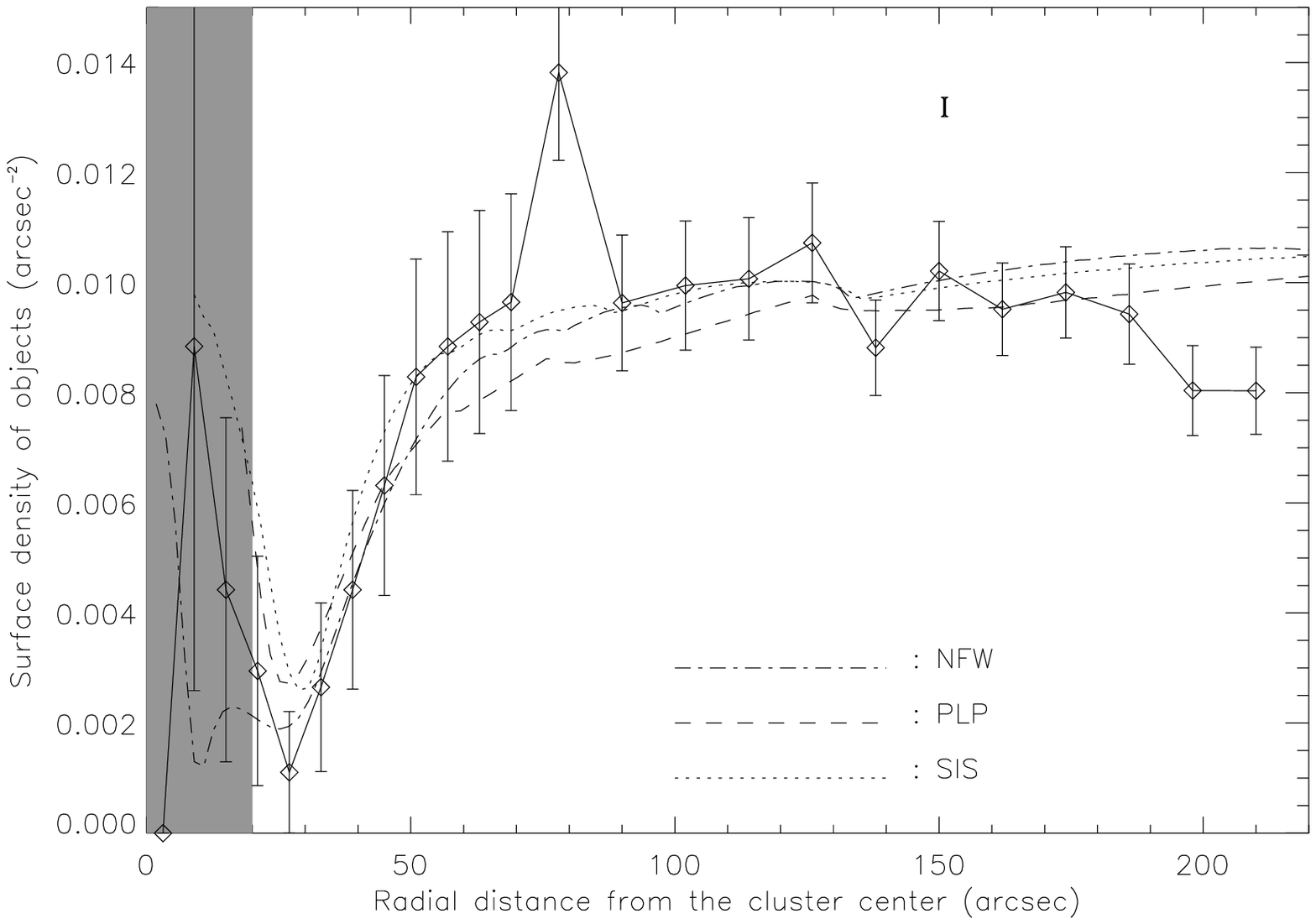}
\includegraphics{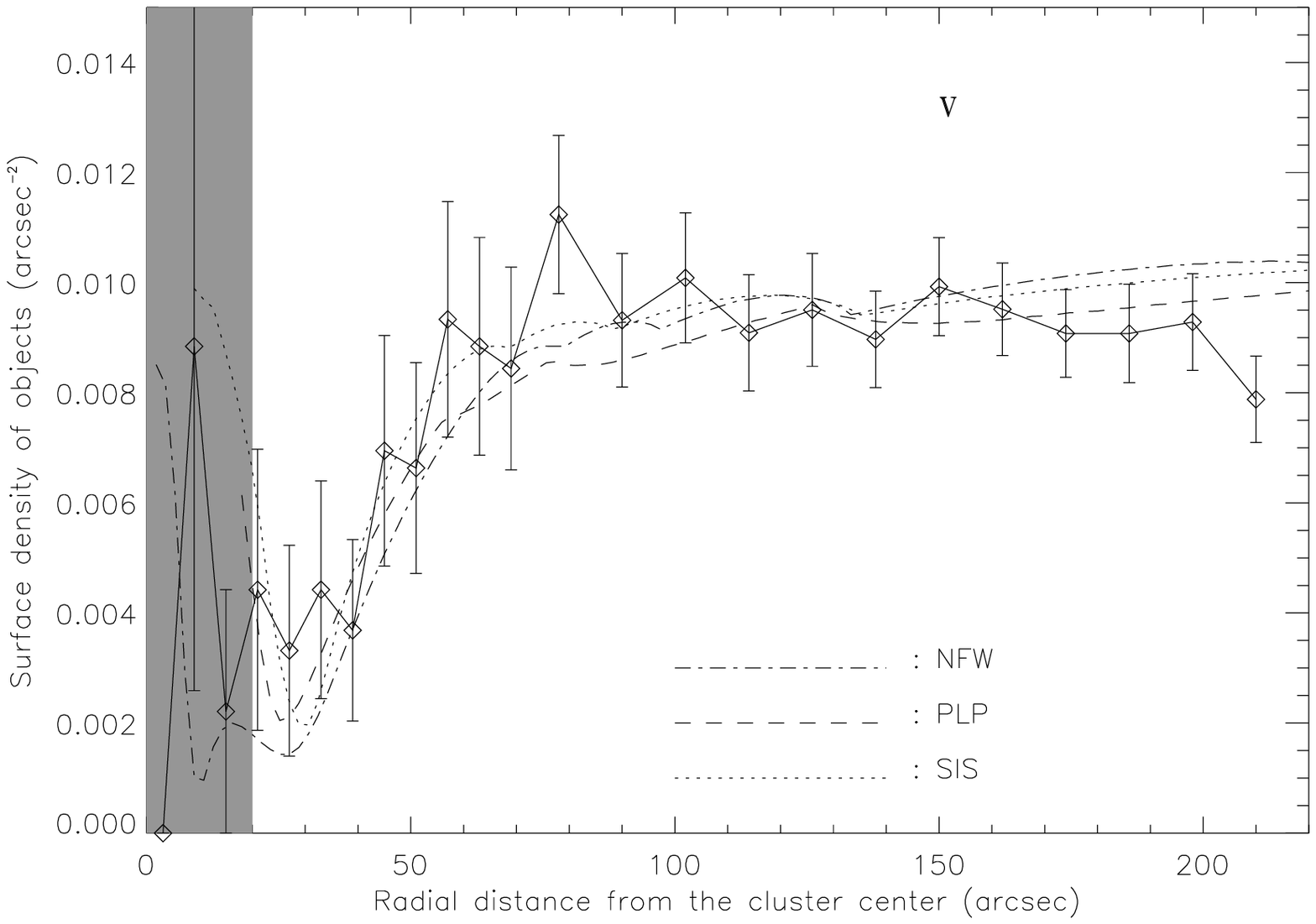}}}
\vspace*{0.5cm}
\centerline{
\resizebox{\hsize}{!}{
\includegraphics{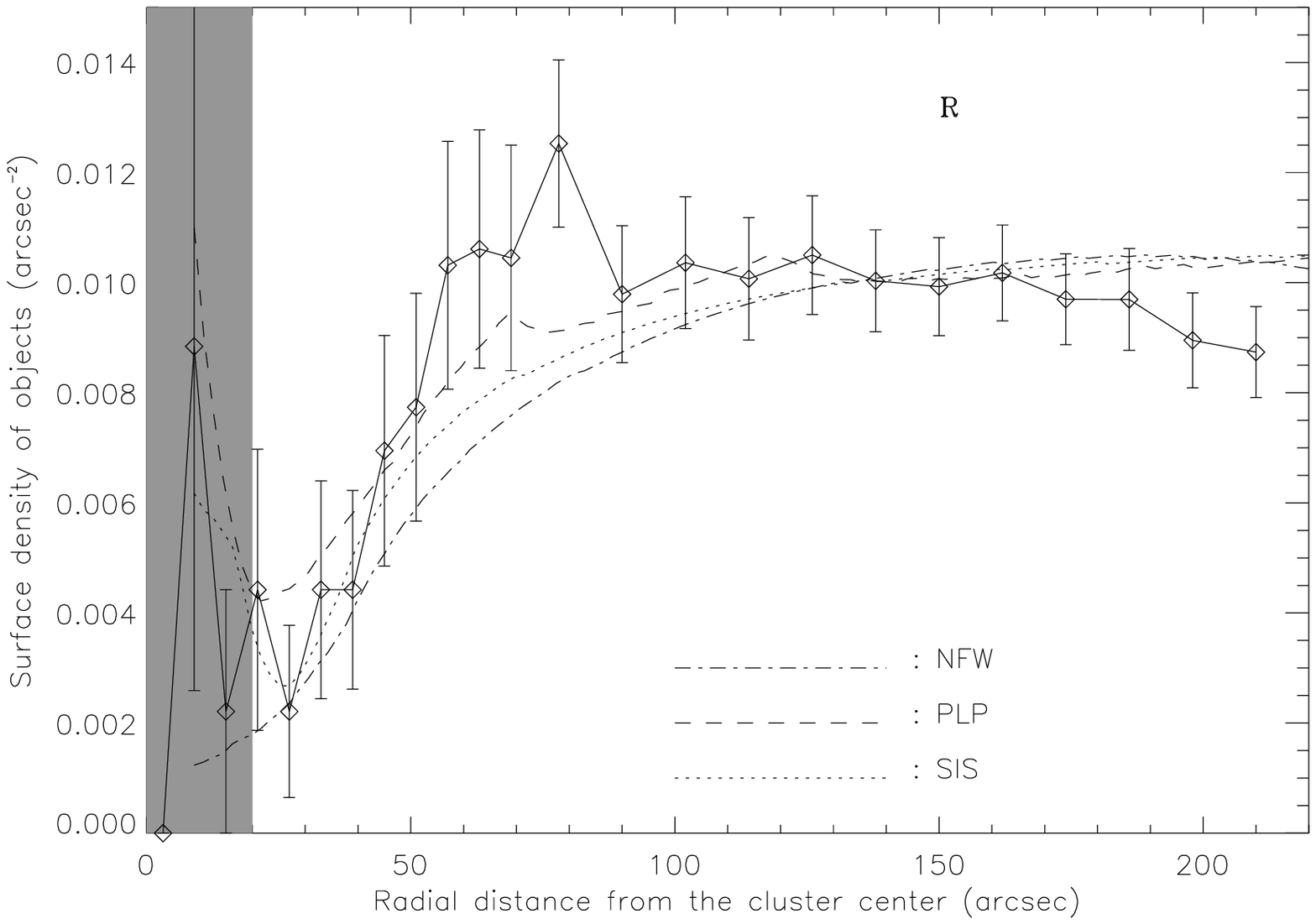}
\includegraphics{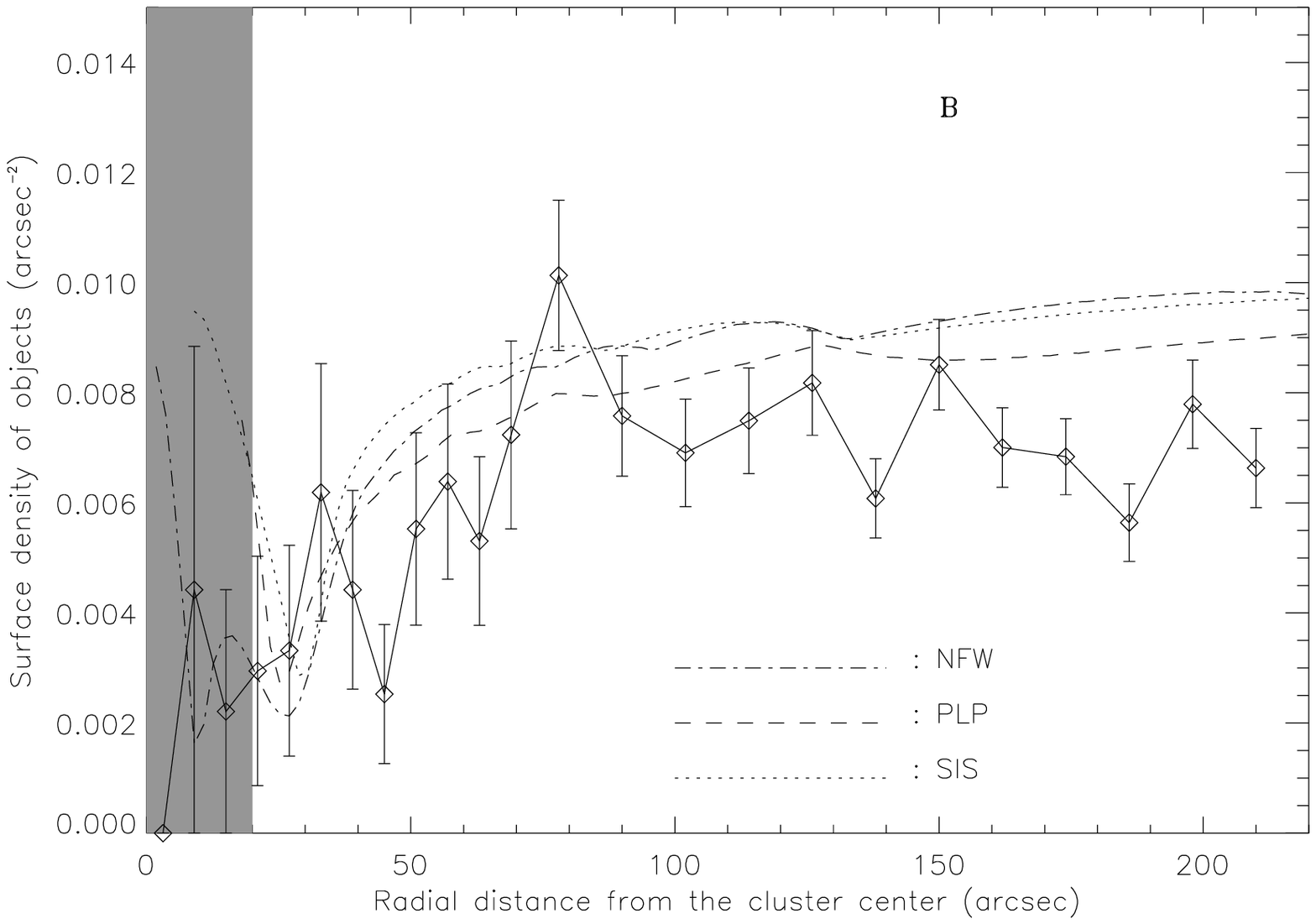}}}
\caption{Depletion curves measured in MS1008--1224 in several spectral 
bands up to the completeness magnitude of the catalogues. For each
plot, the best fit by the three models described in text are
given. Error bars correspond to Poisson statistical noise. The shaded area on 
each plot corresponds to regions where the study of the depletion does not make sense (see text for more details).}
\label{depms1008}
\end{figure*}

The first step would be to locate the cluster center. Its position 
is rather difficult to estimate directly from the
distribution of the number density of background sources, although in
principle one should be able to identify it as the barycenter of the
points with the lower density around the cluster. We did not
attempt to fit it and preferred to fix it 15\arcsec\
North of the cD, following both the X-ray center position 
\cite{lewis} or the weak lensing center \cite{athreya}. 

The radial counts were performed in a range of 3 magnitudes up to the
completeness magnitude in the B, V, R and I bands and up to a radial
distance of 210\arcsec\ from the center, which covers the entire FORS
field. The count step was fixed to 30 pixels (6\arcsec) in the
innermost 80\arcsec\ and for the rest of the field we adopted a count
step of 60 pixels (12\arcsec) to reduce the statistical error
bars. These values are a good balance between statistical errors in
each bin which increase for small steps and a reasonable spatial
resolution in the radial curve, limited by the bin size. 
The depletion curves obtained in the B, V, R and I bands are shown on
Fig. \ref{depms1008}. At some small limiting radius ($r<20\arcsec$), the ratio between 
the area of the count rings and the area covered by the cluster galaxies is close to 
unity. For this reason, we exclude the measures obtained in these innermost rings for any study 
of depletion effects (shaded regions on figure \ref{depms1008}). 

We fitted the observed depletion curve in the I-band with three of
the mass models discussed above, namely a singular isothermal
sphere, a power-law density profile and a NFW profile, and with the galaxy
distribution described in \S \ref{countmodel}. For each model, a 
$\chi^2$ minimization was introduced to derive the best fits and
their related parameters (Table \ref{chi2table}). The first fit
included all the data points ($\chi^2_{\textrm{ap}}$) and gave poor
results with a reduced $\chi^2$ always close to or larger than 2. A second fit was 
done after removing some clear deviant points
($\chi^2_{\textrm{wd}}$): the 
last two points probably poorly corrected from edge effects, and
those associated to the overdensity seen at $r \simeq 80
\arcsec$. This bump is easily identifiable in the V, R and I
curves, and can be partly explained by the presence of a background
cluster lensed by MS1008--1224 and identified by Athreya et al. 
\cite*{athreya}. Nevertheless, even if we remove from our data all the
lower right quadrant of the field where this structure is located, the
bump is still there although significantly reduced. This suggests that
it may be more extended behind the cluster center than initially
suspected. This second fit gave more satisfying results. So we will
consider the parameters associated with the best fit of each model as
our best results.
\begin{itemize}
\item[$\bullet$] The velocity dispersion derived from the SIS model 
($\sigma_{\textrm{fit}} = 1200^{+200}_{-175}$ km s$^{-1}$) is in good
agreement with the value measured by Carlberg et
al. \cite*{carlberg} ($\sigma_{\textrm{obs}} = 1054 \pm 107$ km s$^{-1}$) but is 
more discordant with the value of 900 km s$^{-1}$ inferred from the
shear analysis of Athreya et al. \cite*{athreya}. 
\item[$\bullet$] The slope of the potential fit with a power-law
density profile is close to an isothermal one ($\alpha = 1.88$), 
although slightly shallower. 
\item[$\bullet$] For a NFW profile, we find a virial radius
($r_{200}=3.2 h_{50}^{-1}$ Mpc) and a concentration parameter
($c=8.9$) quite in good agreement with those of Athreya et al.  
\cite*{athreya} derived from weak lensing measures. 
\item[$\bullet$] The comparison between the 3 fits favors a NFW
profile as the best fit of our depletion curves (Table
\ref{chi2table}), also in agreement with the shear results for this
cluster.
\end{itemize}

\begin{table}
\caption{Results of the fit of the model parameters with a $\chi^2$
minimization. The errors are given at the 99.9 \%\ confidence level.}
\label{chi2table}
\begin{center}
\begin{tabular}{ccc}
\hline\noalign{\smallskip}
Model & Parameter & reduced $\chi^2$ \\
\noalign{\smallskip}\hline \\
\noalign{All points \hfill} \\
SIS & $\sigma=1200\pm^{100}_{75}$ km s$^{-1}$ & 1.85 \\
PLP & $\rho_E=2.4\pm^{0.5}_{0.4} \times 10^{15}$ M$_{\sun}/$Mpc$^3$ 
& 1.87 \\
PLP & $\alpha=1.88\pm^{0.08}_{0.12}$ & 1.89 \\
NFW & $r_{200}=3.2\pm^{0.2}_{0.4}$ Mpc & 2.16 \\
NFW & $c=8.9\pm^{2.3}_{2.0}$ & 2.18 \\
\\
\noalign{Without 3 deviant points \hfill} \\
SIS & $\sigma=1200\pm^{200}_{175}$ km s$^{-1}$ & 0.49 \\
PLP & $\rho_E=2.4\pm^{1.3}_{0.6} \times 10^{15}$ M$_{\sun}/$Mpc$^3$ 
& 0.62 \\
PLP & $\alpha=1.88\pm^{0.27}_{0.25}$ & 0.65 \\
NFW & $r_{200}=3.2\pm^{0.7}_{0.6}$ Mpc & 0.44 \\
NFW & $c=8.9\pm^{6.7}_{4.0}$ & 0.45 \\
\noalign{\smallskip}\hline
\end{tabular}
\end{center}
\end{table}

\begin{figure}
\centerline{
\resizebox{\hsize}{!}{
\includegraphics{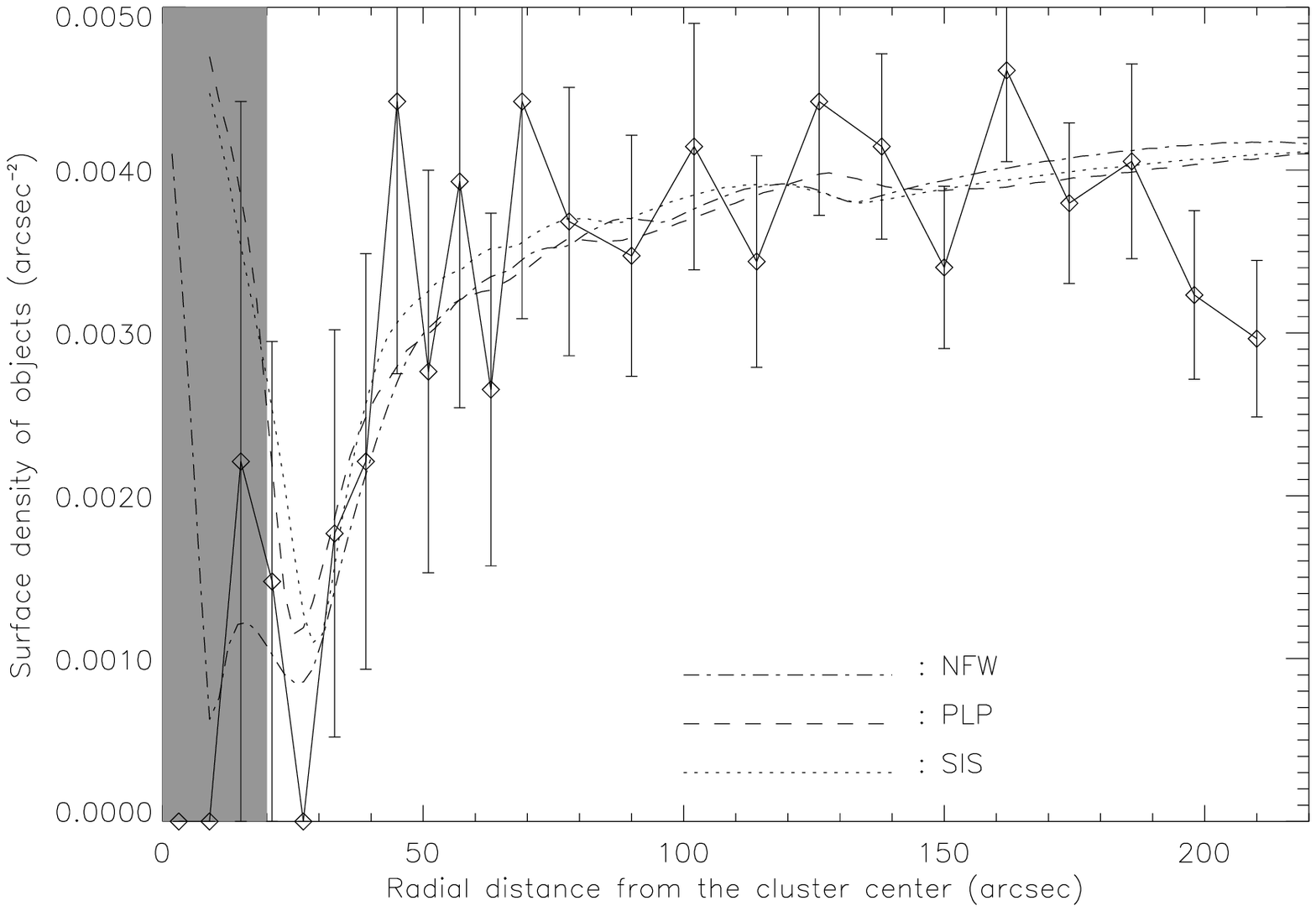}}}
\centerline{
\resizebox{\hsize}{!}{
\includegraphics{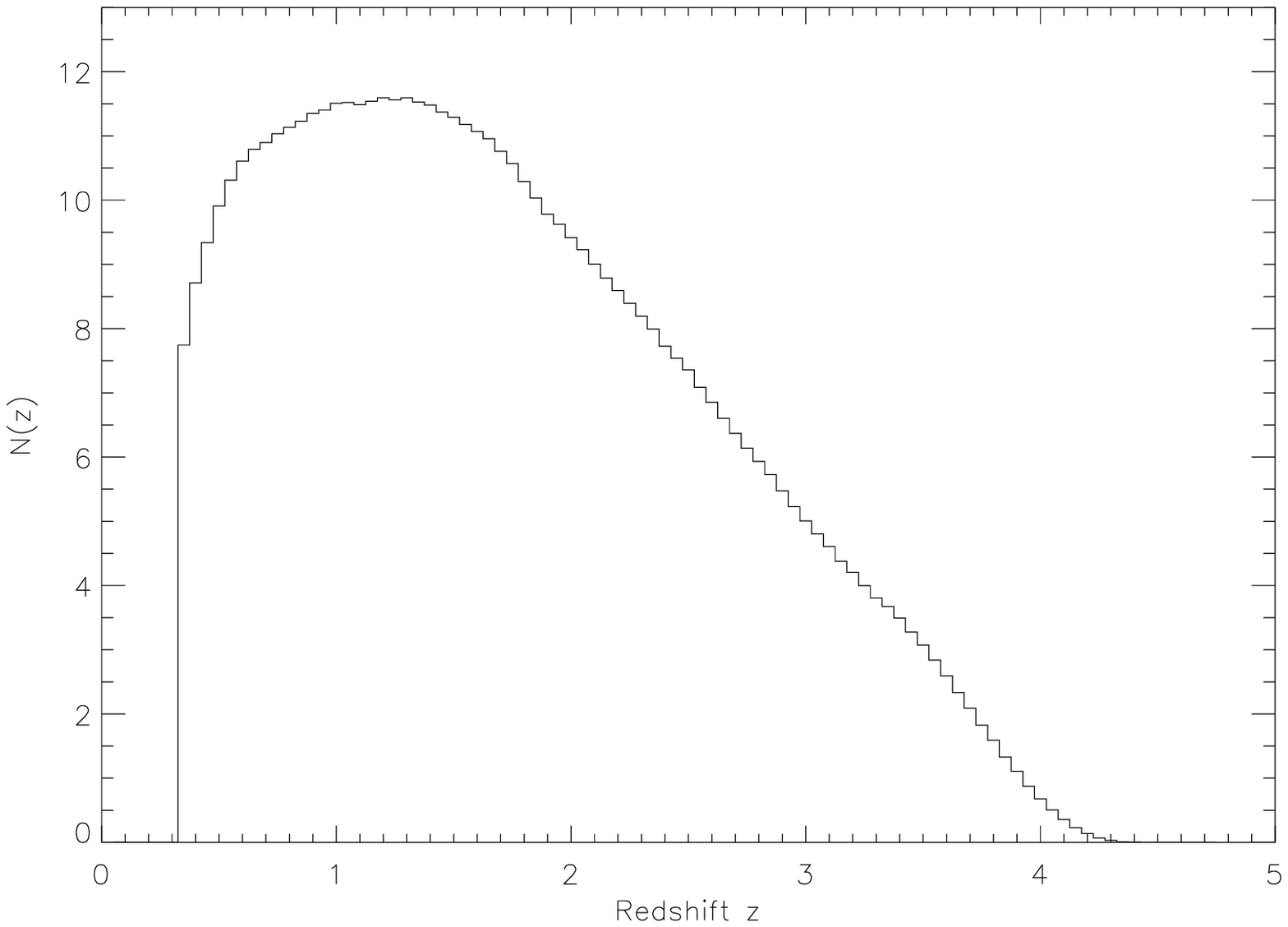}}}
\caption{{\bf Top: } Depletion curve obtained for $26<B<28$. The fits by the
three models described in text are given. {\bf Bottom: } Redshift
distribution of background objects which corresponds to the best fit
of the above depletion curve.}
\label{ms1008deepB}
\end{figure}

Note however that we did not use the B-band data in the fit because
the depletion effect is less obvious and much noisier than in the 3
other filters. This is due to the fact that for this magnitude
range the logarithmic slope of the counts is close to the critical
value 0.4 and the depletion effect is strongly attenuated. Probing
fainter objects in this band ($26<\mathrm{B}<28$), even beyond the
completeness limit, strengthens the depletion signal again , as it is
shown in Figure \ref{ms1008deepB}. In addition this deeper magnitude
range may be used to scan a different redshift distribution with a
higher density of high redshift objects. This is one of the future
prospects of the depletion curve analysis to perform a
multi-wavelength analysis to try to disentangle lensing effects from
the statistical distribution of the sources. 

\subsection{Ellipticity and orientation of the mass distribution}
\begin{figure}
\centerline{
\resizebox{\hsize}{!}{
\includegraphics{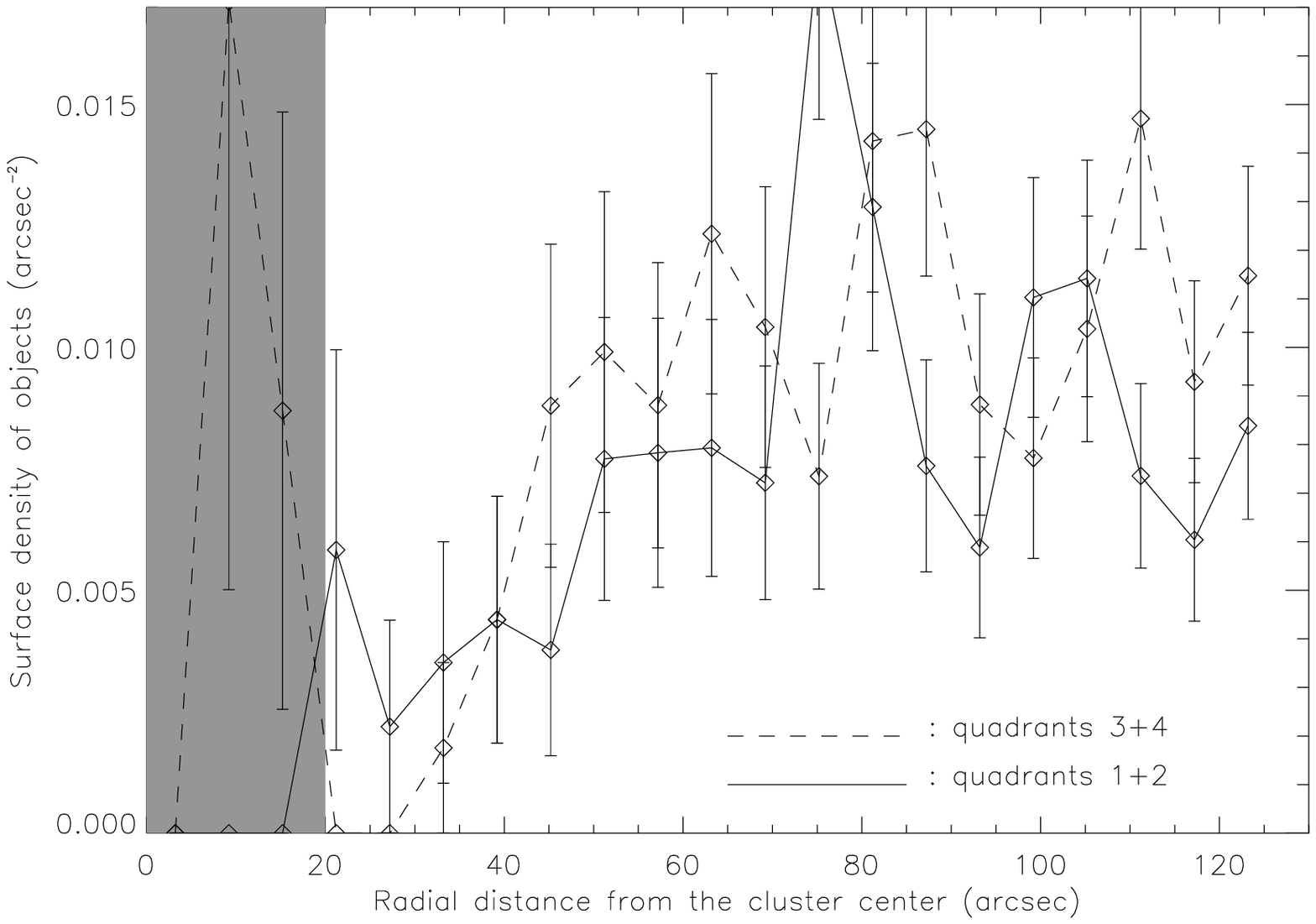}}}
\centerline{
\resizebox{\hsize}{!}{
\includegraphics{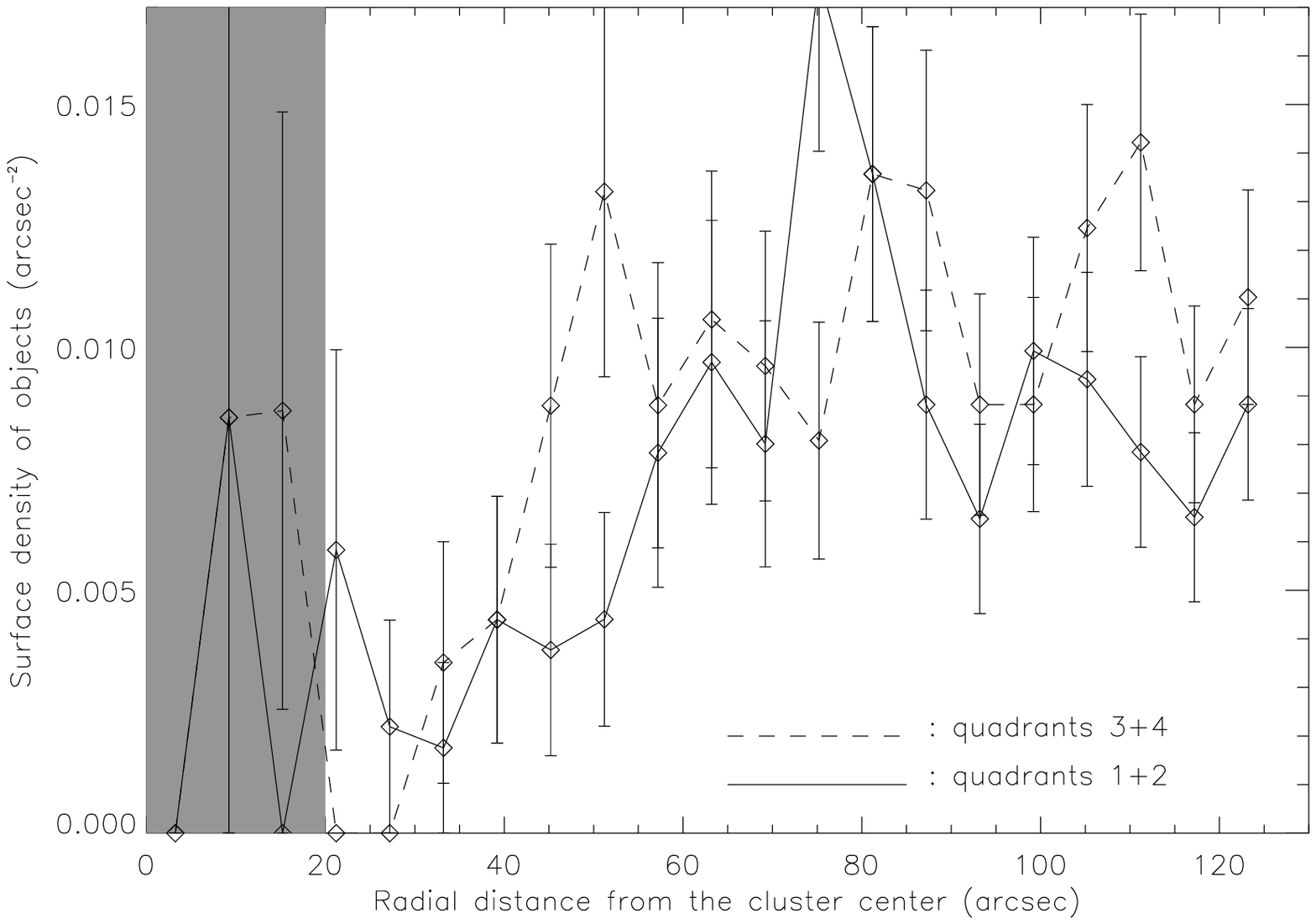}}}
\centerline{
\resizebox{\hsize}{!}{
\includegraphics{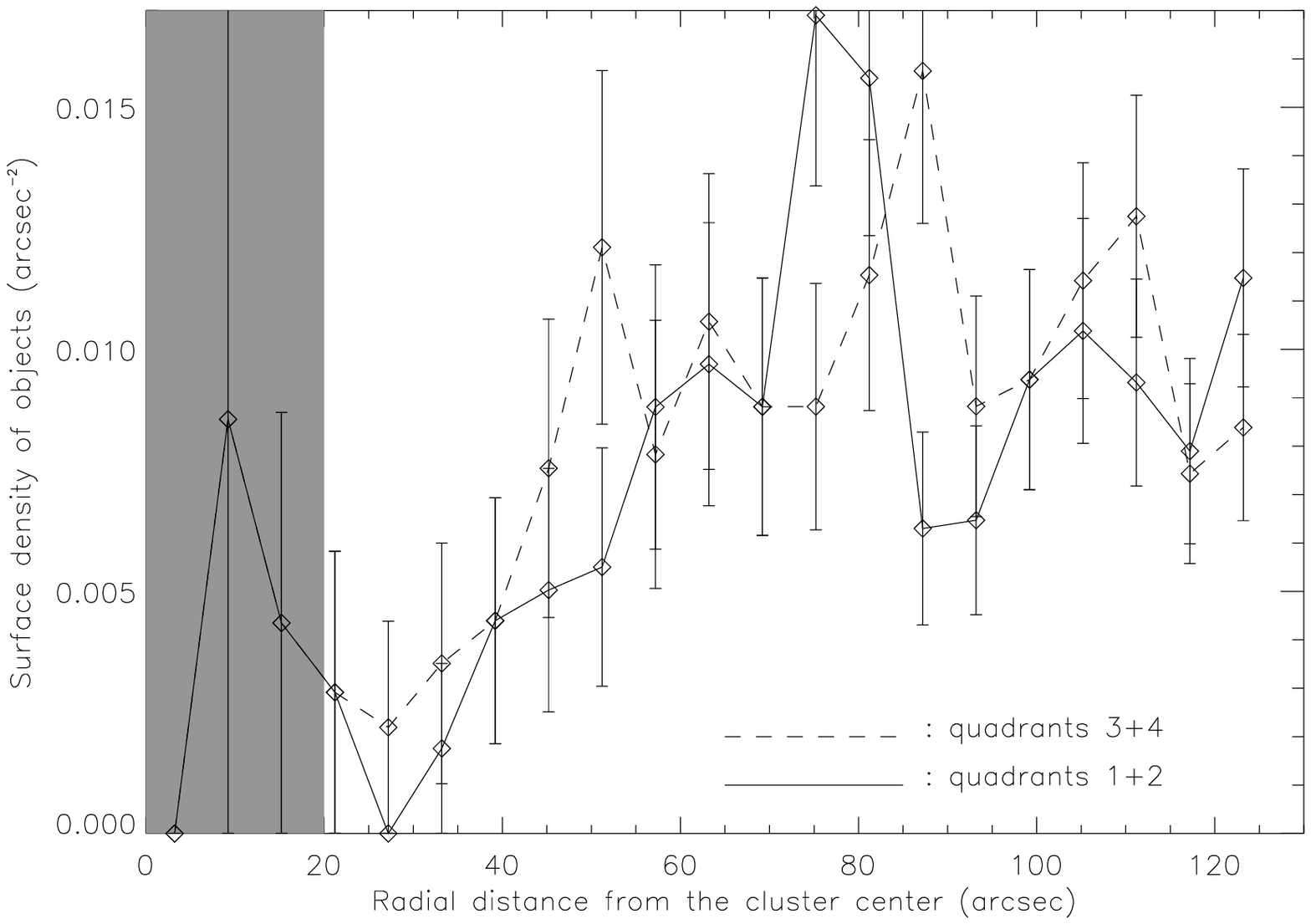}}}
\caption{Depletion curves obtained in the I band  
for $\theta=0^{\circ}$ (top), $\theta=-10^{\circ}$ (middle) and
$\theta=-20^{\circ}$ (bottom). The radial step has been reduced to
6\arcsec\ only for a more accurate localisation of the minima}
\label{epsilonms1008}
\end{figure}
We have also used the depletion effect to determine the orientation of
the main axis and the ellipticity of MS1008--1224. For this purpose,
we divided the area of the FORS field into four quadrants where
quadrants 1 and 2 are in opposition, and we performed the counts for
radial distances from the cluster center to 130\arcsec\
(Fig. \ref{ms1008B}). The positions of the four quadrants are marked
by $\theta$, the clockwise angle between the North-South axis and the
median of quadrant 1. We then computed the depletion curves
corresponding to the counts in quadrants 1+2 and 3+4. We varied
$\theta$ and followed the shift of the minimum position of the
depletion area of the two curves.  We can measure the orientation of
the two main axis when the shift between the two minima is
maximum. For this value of $\theta$, the relative positions of the two
minima give the main axis ratio and consequently the ellipticity. The
results, obtained from I-band data only, are similar in the others
bands. From the I-band data only, we found $\theta=-10^{\circ} \pm
3^{\circ}$ and an ellipticity of $\epsilon=0.18
\pm 0.04$ (Fig. \ref{epsilonms1008}). These results agree well with
the mass map orientation derived from weak lensing as well as the
galaxy number density map \cite{athreya}. A small discrepancy arrises 
with the X-ray gas distribution which orientation is shifted a few 10
to 20 degrees West from our measure \cite{lewis}. But globally, all
these results are consistent within each other.


\section{Conclusions}

The gravitational magnification by cluster lenses represents 
an effective tool to probe the distant universe and the mass
distribution in the lenses. The first aim of this paper was to study the
effects of lens mass profile model parameters on the typical features
of depletion curves. We have also attempted to characterize some
features associated with the background redshift distribution of the
galaxies. Models were constructed in three bands covering a large
spectral range and by using five different lens models.

Our simulations agree well with very deep and high quality
images of the cluster MS1008--1224 obtained with the VLT and FORS. The
depletion effect is clearly seen in this cluster, and we have fitted 
its radial variation with several sets of mass profiles. The results
are quite satisfying as we are able to constrain the mass profile up
to a reasonable distance from the center (about 200\arcsec, or
equivalently 1.1 $h_{50}^{-1}$ Mpc). Our results marginally favor the 
NFW mass profile over the isothermal profile, and more significantly
reject a power-law distribution, essentially because of the steep
rise of the depletion curve just after its minimum. Note that this
region corresponds to the ``intermediate'' lensing regime, where shear
measurements are more difficult to relate to the mass distribution. We
have also studied the shape of the depletion area, which is easy to
relate to the ellipticity of the mass distribution. We were then able
to constrain the ellipticity and orientation of the potential of
MS1008--1224 with a good accuracy.

This preliminary study highlights the need for additional exploration 
of several issues not fully explored in the present paper. For 
example, the question of clustering of the background sources
still remains, although in the case of MS1008--1224, we have shown how
it can be partly eliminated. Schneider et al. (2000) also mention
this problem and insist on the fact that, at deep magnitudes, the two
point correlation function of the sources is still quite uncertain,
but a positive signal does not seem to extend much above $\sim$ 1 
\arcsec. More quantitatively, we can use the most recent 
measures of the two point correlation function from the HDF-South
\cite{fynbo}. At the magnitude limits used in our paper, this
correlation function does not exceed a few percent for an angular
separation of 10\arcsec , typical of our bin size. Previous
measurements were less optimistic in the sense that their values attained 
8 to 10 \%\ at 10\arcsec , even at very faint
magnitudes. But although there is still only poor knowledge of the
amplitude of the correlation function at the faintest levels, we can
hope that the effect will not be dramatic in our studies of the
magnification bias, provided we retain a reasonable bin size at 
several arcseconds, to wash out most of the inhomogeneities.

Following preliminary approaches from the observational point of
view \cite{taylor,athreya} or a more theoretical one
\cite{schneiderking}, one now clearly needs to extensively and
quantitatively compare the weak lensing approach with the
magnification bias. In particular, with the new facilities of deep
wide-field imaging presently available, most of the difficulties
related to a small field-of-view can be overcome: for weak lensing
measurement, a complete mass reconstruction requires shear measurements
up to the ``no-shear'' region in the outer parts of the cluster to
integrate the mass inwards. The absolute normalization of the field
number counts for the magnification bias can also be estimated outside
the cluster, in exactly the same observing conditions (filter,
magnitude limit, seeing, \ldots ), giving an absolute calibration of
the depletion effect.

Moreover, the full 2D mass reconstruction of the cluster from the
depletion signal alone should be tested. As the signal is directly
related to the magnification $\mu ( \vec r )$, and then to $\kappa (
\vec r )$ and $\gamma ( \vec r )$, it should be in principle feasible
to invert the depletion map to produce a non-parametric mass map. In
practice, the reconstruction is simpler as soon as the shear $\gamma$
becomes small, because in that case $\mu$ and $\kappa$ are simply 
linearly related. In any case, we have shown in this paper that it is
rather easy to reach the outer parts of the cluster up to Mpc
scales. This is quite similar to what can be done with deep X-ray maps
for intermediate redshift clusters \cite{soucail}. We want to
insist on the fact that the magnification bias effect is easy to
detect from the observational point of view because it is less
sensitive to seeing conditions or to geometrical distortions of the
instruments than are shear measurements \cite{broadhurst2,fort}. It only requires deeper
observations, not necessarily in photometric conditions, provided one
is able to reach the outer parts of the cluster to normalise the
number counts. In order to improve the mass reconstruction, one may
progress with the help of photometric redshifts, quite useful for
faint objects, to constrain the redshift distribution of the sources
\cite{pello,bolzonella}. This approach might also be useful 
in order to limit the influence of large 
scale over-density fluctuations when evaluating the magnification and the asymptotic 
limit of the depletion curves. This of course requires deep multi-color
photometry, such as the one obtained on MS1008--1224. Extending this
study up to large-scale structures (LSS) would probably be more
difficult to implement as compared to the search for cosmic shear
\cite{vanwaerbeke}, and the depletion effect should be restricted to
cluster scales, at least with the simple method used in this paper. 

Finally, one of our initial prospects was to try to constrain the
background redshift distribution with multi-wavelength observations of 
the depletion effect. We have shown that this is quite a difficult
task as the wavelength dependence of the depletion curves is a kind of 
second order effect. Nevertheless, this may be an interesting point to explore
at other wavelengths, where the background redshift distribution is
quite different than in the optical. For example, deep ISO
observations in the mid-IR of a few cluster lenses \cite{altieri} may 
represent an extension of our analysis, as well as submm observations with SCUBA, 
provided enough sources are detected behind the lenses for a
statistical analysis. Another possibility would be to address the
question of the nature of the X-ray background sources and their
redshift distribution through deep and high resolution observations of 
clusters with the new X-ray satellites Chandra and XMM \cite{refregier}.

\acknowledgements We wish to thank Roser Pell\'o and Bernard Fort for
fruitful discussions and encouragements. This work was supported by
the TMR network ``Gravitational Lensing : New Constraints on Cosmology
and the Distribution of Dark Matter'' of the European Commission under
contract No : ERBFMRX-CT98-0172.
 


\end{document}